%% file: main.tex
\documentclass[lettersize,journal]{IEEEtran}

\ifCLASSOPTIONcompsoc
  \usepackage[nocompress]{cite}
\else
  \usepackage{cite}
\fi

\usepackage{tcolorbox}
\usepackage{amsfonts} 
\usepackage{amssymb}
\usepackage{array}
\usepackage{lipsum}
\usepackage{subcaption}
\usepackage{longtable}
\usepackage{xspace}
\usepackage{algorithm}
\usepackage[noend]{algpseudocode}
\usepackage{amsmath}

\usepackage{subcaption}
\usepackage{relsize}
\usepackage{xcolor}
\usepackage{mathtools}
\usepackage{multirow}
\usepackage{marginnote}
\usepackage{pifont}
\usepackage{url}

\newcolumntype{L}[1]{>{\raggedright\let\newline\\\arraybackslash\hspace{0pt}}m{#1}}
\newcolumntype{C}[1]{>{\centering\let\newline\\\arraybackslash\hspace{0pt}}m{#1}}
\newcolumntype{R}[1]{>{\raggedleft\let\newline\\\arraybackslash\hspace{0pt}}m{#1}}

\newcommand{\mIoU}{$\mathit{mIoU}$\xspace}

\newcommand{\UMARS}{\MARSSIM\xspace}

\newcommand{\DESIGNATE}{DESIGNATE\xspace}
\newcommand{\VDA}[1]{$\mathbf{A_{12}}$}

\newcommand{\APPR}{ORBIT\xspace} 
\newcommand{\APPRFLIP}{ORBIT$_{flip}$\xspace}
\newcommand{\APPRNOISE}{ORBIT$_{noise}$\xspace} 
\newcommand{\APPRSURPRISE}{ORBIT$_{SA}$\xspace} 
\newcommand{\APPRUNCERTAIN}{ORBIT$_{MCD}$\xspace} 
\newcommand{\FLIP}{$_{flip}$\xspace}
\newcommand{\NOISE}{$_{noise}$\xspace} 
\newcommand{\SURPRISE}{$_{SA}$\xspace} 
\newcommand{\UNCERTAIN}{$_{MCD}$\xspace} 
\newcommand{\APPRFLIPCYCLE}{ORBIT$_{flip,cycleGAN}$\xspace}
\newcommand{\APPRNOISECYCLE}{ORBIT$_{noise,cycleGAN}$\xspace} 
\newcommand{\APPRSURPRISECYCLE}{ORBIT$_{SA,cycleGAN}$\xspace} 
\newcommand{\APPRUNCERTAINCYCLE}{ORBIT$_{MCD,cycleGAN}$\xspace} 
\newcommand{\FLIPCYCLE}{$_{flip,cycleGAN}$\xspace}
\newcommand{\NOISECYCLE}{$_{noise,cycleGAN}$\xspace} 
\newcommand{\SURPRISECYCLE}{$_{SA,cycleGAN}$\xspace} 
\newcommand{\UNCERTAINCYCLE}{$_{MCD,cycleGAN}$\xspace}

\definecolor{mygreen}{rgb}{0.0, 0.2, 0.13}

\newcommand{\MARSSIM}{MarsSim\xspace}

\begin{document}

\title{GAN-enhanced Simulation-driven DNN Testing in Absence of Ground Truth}

\author{Mohammed~Oualid~Attaoui and Fabrizio~Pastore

\IEEEcompsocitemizethanks{\IEEEcompsocthanksitem M. O. Attaoui and F. Pastore are with the SnT Centre, University of Luxembourg.\protect\\
E-mail: mohammed.attaoui@uni.lu, fabrizio.pastore@uni.lu
}
}

\IEEEtitleabstractindextext{%
\begin{abstract}
The generation of synthetic inputs via simulators driven by search algorithms is essential for cost-effective testing of Deep Neural Network (DNN) components for safety-critical systems. However, in many applications, simulators are unable to produce the ground-truth data needed for automated test oracles and to guide the search process.

To tackle this issue, we propose an approach for the generation of inputs for computer vision DNNs that integrates a generative network to ensure simulator fidelity and employs heuristic-based search fitnesses that leverage transformation consistency, noise resistance, surprise adequacy, and uncertainty estimation. We compare the performance of our fitnesses with that of a traditional fitness function leveraging ground truth; further, we assess how the integration of a GAN not leveraging the ground truth impacts on test and retraining effectiveness. 

Our results suggest that leveraging transformation consistency is the best option to generate inputs for both DNN testing and retraining; it maximizes input diversity, spots the inputs leading to worse DNN performance, and leads to best DNN performance after retraining.
Besides enabling simulator-based testing in the absence of ground truth, our findings pave the way for testing solutions that replace costly simulators with diffusion and large language models, which might be more affordable than simulators, but cannot generate ground-truth data.
\end{abstract}

}

\maketitle

\IEEEdisplaynontitleabstractindextext

\IEEEpeerreviewmaketitle

\input{introduction.tex}
\input{background.tex}
\input{approach.tex}
\input{evaluation.tex}
\input{related.tex}

\input{conclusion.tex}

\section*{Acknowledgments}
The experiments presented in this paper were carried out using the HPC facilities of the University of Luxembourg (see \url{http://hpc.uni.lu}). This work has been partially supported by the ESA contract RFP/3-17931/22/NL/GLC/my,
TIA (Test, Improve, Assure). 
\bibliographystyle{plain}
\bibliography{bibliography}

\end{document}

%% file: introduction.tex
\section{Introduction}
\label{sec:intro}
In many safety-critical, cyber-physical systems, Deep Neural Networks (DNNs) are becoming essential to automate tasks that are challenging to program manually, such as vision-based autonomous navigation in drones, satellites, robots, and rovers~\cite{DEEPVO,SpaceDNN,Space2,Pugliatti2022,REN2021329,nerf-nav}.
Despite their effectiveness, the vast input space complicates the validation of DNNs across all relevant usage scenarios, which is necessary for safety-critical systems~\cite{SOTIF}. Indeed, this validation becomes impractical when real-world data collection is prohibitively expensive or unfeasible, as in space applications~
\cite{Azzalini2023,Zysk2023}.


Simulators can be effectively combined with search algorithms to detect scenarios leading to DNN failures~\cite{10.1145/3510003.3510188,Hazem:SEDE,Gambi2019}. However, the fidelity gap between simulator outputs and real-world data undermines the reliability of testing results. This challenge has recently driven  the adoption of Generative Adversarial Networks (GAN~\cite{GANs}) to enhance simulator outputs and generate realistic images~\cite{stocco2022mind,Amini2024,attaoui2024designate}. For instance, we developed \DESIGNATE, an approach utilizing the Pix2PixHD~\cite{wang2018pix2pixHD} GAN to produce realistic images for computer vision tasks. 
\DESIGNATE has been successfully employed to evaluate and enhance DNNs processing Martian and urban landscapes.


Fidelity alone is insufficient for the successful validation of DNN-based components. Simulators must also provide ground-truth data to enable the definition of automated oracles. For example, autonomous driving simulators offer middle-lane positions, allowing the determination of whether an autonomously driven car goes off-the-lane. Unfortunately, simulators often do not provide the necessary ground truth for oracle automation. Specifically, while market-available simulators can generate data for various DNN-automated tasks, they typically provide ground-truth data only for the task that motivated their development.
For example, flight simulators~\cite{flightgear,gazebo} can produce image sequences useful for testing vision-based autonomous flight control systems and DNNs for soil elevation prediction~\cite{elevation}. However, they do not generate slope and heat maps, thus preventing the assessment of soil elevation prediction. Similarly, popular simulators for autonomous driving tasks, such as Carla~\cite{carla}, AirSim~\cite{airsim}, and BeamNG~\cite{beamng}, cannot generate slope and heat maps. Furthermore, space simulators such as PANGU~\cite{PANGU} and Surrender~\cite{SurRender2025} cannot produce segmentation labels for object identification, thus preventing the testing and retraining of DNNs for vision-based navigation.
Extending simulators to generate the required ground truth is generally too expensive for companies developing DNNs. For example, space navigation components might be developed by small startups that can't afford the development of dedicated simulators.


The absence of ground truth is a problem that is likely to become more pronounced in the near future, when simulators will be combined with diffusion~\cite{diffusion} and large language models~\cite{LLMs} to generate test data~\cite{Baresi2025}. For example, it is infeasible to leverage a diffusion model to generate both a landscape image and a corresponding segmentation map (i.e., an image where pixels' color depends on the type of object they belong to).

To enable testing and improvement of DNNs in absence of ground truth, we propose Oracle-based Iterative Testing (\APPR), a search-based technique that drives input generation by leveraging approaches to implement DNN testing oracles in absence of ground-truth. \APPR extends  DESIGNATE by eliminating the need for ground truth to compute fitness. Specifically, we propose four  fitness functions that leverage transformation consistency (flipping), noise resistance, activation-based adequacy (surprise adequacy~\cite{KimSurprise2023}), and uncertainty estimation (Monte Carlo Dropout~\cite{gal2016dropout}). 
By leveraging these fitness functions, \APPR guides test input generation through model behavior rather than explicit correctness labels. Further, instead of leveraging a GAN that requires the ground truth to generate realistic images, it leverages a GAN that generates realistic images directly from simulator images (i.e., CycleGAN~\cite{zhu2017unpaired}).


We assessed \APPR with a subject DNN for the segmentation of Mars images that was developed within a project with the European Space Agency (ESA~\cite{ESA}).
Our fitness functions perform similar to a ground-truth-based fitness function for testing effectiveness; they lead to failures of comparable severity and generate images as diverse as those produced by \DESIGNATE (with negligible effect size). 
Additionally, when the generated images are used to retrain the DNN, they all result in better performance than a random approach, with the flipping-based fitness function achieving the same effectiveness as \DESIGNATE. Last, we demonstrated that GANs that do not require ground truth still explore the same areas of the input space during testing, albeit with different frequencies, while achieving better performance improvements than GANs that require ground truth.

Since our results indicate that the proposed fitness functions effectively address the lack of ground truth, we believe they could be integrated not only in simulator-driven testing approaches (our work) but also in approaches leveraging diffusion models~\cite{Baresi2025,Baresi2025b} (future work).

This paper proceeds as follows. Section~\ref{sec:background} introduces background work, Section~\ref{sec:approach} describes \APPR, Section~\ref{sec:evaluation} describes our empirical evaluation, Section~\ref{sec:related} summarizes related work,
Section~\ref{sec:conclusion} concludes the paper.

%% file: background.tex
\section{Background}
\label{sec:background}

\subsection{Semantic Segmentation}
\label{sec:background:segmentation}

As running example, in this paper, we consider DNNs performing semantic segmentation of Mars landscape images.
Semantic segmentation is a computer vision task that assigns a class label to every pixel in an image. Unlike other image recognition or classification methods, it offers a detailed understanding of an image's content by labeling each pixel with a specific class, such as ``bedrock'', ``sand'', ``big rock'', ``small rock'', ``soil'', ``other'' (e.g., sky)~\cite{ai4mars2020}.


The accuracy of one semantic segmentation prediction can be computed with the  \emph{mean Intersection over Union (mIoU)} metric. The \emph{Intersection over Union (IoU)} metric, also called Jaccard Index~\cite{everingham2010pascal}, measures the overlap between the predicted segmentation mask and the ground truth mask for a class (e.g., ``bedrocks''). 
The $mIoU$ for one image is thus obtained by averaging the IoU scores across all classes of interest:
\begin{equation}
mIoU = \frac{1}{C} \sum_{c=1}^{C} \frac{TP_c}{TP_c + FP_c + FN_c}
\end{equation}
with $TP_c$, $FP_c$, and $FN_c$ being the number of image pixels that are correctly assigned, erroneously assigned, and erroneously not assigned to class $c$, respectively.

\subsection{Simulation environments}
\label{sec:background:simulator}
Simulators are crucial for DNN testing and development, especially when real-world data collection is impractical. 
However, because of the complex characteristics of the environment to be simulated, simulators are often specialized and can hardly be extended for other purposes, which motivates our work.

In the space context, PANGU (Planet and Asteroid Natural Scene Generation Utility)~\cite{PANGU,PANGUtr,PANGUpaper} is a high-fidelity simulation tool, considered the state-of-the-art in rendering celestial bodies. 
It is widely used by space companies including ESA. 
PANGU includes a wide range of features such as high-resolution surface models (i.e., accurate representations of planetary surfaces based on topographic data), sensor simulation (cameras, lidar, and radar, including noise caused by optical distortion, rolling shutter, radiation), generation of scenes with moving elements (i.e., rover navigation and object manipulation). Unfortunately, despite its comprehensive set of features, PANGU cannot directly generate segmentation labels, requiring costly development and integration of additional simulation models.
Other well-known space simulators can't be leveraged for the generation of terrain segmentation data as well. For example, Surrender, which is
a proprietary simulator used to validate vision-based applications involving 
Airbus~\cite{SurRender2025,Lebreton2021} can generate images, radiance, depth map, slope map but not segmentation labels.
Corto~\cite{CORTO} generates synthetic image-label pairs of celestial and artificial bodies to be used for the analysis of trajectories and the segmentation of celestial bodies surface; however, it can't be used to simulate terrain exploration. 

Similar limitations can be observed in simulators for other contexts. For avionics, FlightGear~\cite{flightgear} cannot generate slope, depth, or heat maps and thus can't be used to train DNNs that predict such information by means of computer vision; similarly,
Gazebo~\cite{gazebo} does not generate slope maps. For automotive, the Carla~\cite{carla} and AirSim~\cite{airsim} simulators do not generate slope and heat maps, while
BeamNG~\cite{beamng} does not generate depth, slope, and heat maps.

To conduct our study, we leverage a custom-built Mars simulator (hereafter, \emph{\MARSSIM}) designed to simulate images of the Martian surface and generate corresponding segmentation labels (rocks, bedrocks, soil, and sand). The simulator is built on Unreal Engine~\cite{unrealengine}, leveraging the same plugins used by AirSim~\cite{shah2018airsim} for navigation and semantic segmentation (ground truth generation). Different from PANGU and the space simulators presented above, \MARSSIM does not simulate sensors and object dynamics because they are out-of-scope in this paper. \MARSSIM receives as input the position and orientation of the rover.



\subsection{Generative Adversarial Networks}
\label{sec:background:gan}
Generative Adversarial Networks (GANs) are a class of machine learning models designed to generate realistic data by learning the distribution of a given dataset~\cite{GANs}. 
GANs have demonstrated remarkable success in tasks like image generation, image-to-image translation, and data augmentation. In this study, we use  Pix2PixHD~\cite{wang2018pix2pixHD} and CycleGAN~\cite{zhu2017unpaired}.

Pix2PixHD is a supervised image-to-image translation framework that learns a mapping from input images to output images using paired datasets. The model is based on conditional GANs, where the generator aims to transform input images into a target domain, and the discriminator evaluates the realism of the generated outputs relative to the paired ground truth.
Following our previous work~\cite{attaoui2024designate}, we use Pix2PixHD to transform segmentation masks generated by \MARSSIM into realistic Martian images; this task is enabled by the availability of pairs $\langle$segmentation image, corresponding real-world image$\rangle$ from the AI4MARS dataset. Pix2PixHD cannot be used to transform \MARSSIM images into realistic images because there is no dataset that pairs \MARSSIM images with real Mars landscapes.


CycleGAN is an unsupervised image-to-image translation framework that enables mapping between two domains without requiring paired datasets. CycleGAN introduces a cycle-consistency loss to enforce consistent outputs in translations from one domain to another and back to the original. 
In this study, we use CycleGAN to transform synthetic Mars images into realistic ones, eliminating the need for ground truth segmentation masks.

\subsection{DESIGNATE}
\label{sec:background:designate}
\DESIGNATE \cite{attaoui2024designate} is an approach for testing and retraining vision-based DNNs by combining GANs with meta-heuristic search and simulators. GANs are used to generate images that are realistic (i.e., similar to real-world ones). 
Search is used to drive the similator towards the generation of images that are diverse and lead to worst DNN performance. In previous work~\cite{attaoui2024designate}, we applied \DESIGNATE to both an autonomous driving context (leveraging the AirSim simulator) and Mars segmentation (using \MARSSIM).

The output of DESIGNATE is a set of realistic and diverse images stored in an archive that reveal DNN limitations; they can either (1) be inspected to determine when the DNN may fail or (2) used to automatically retrain the DNN.

Figure \ref{fig:designate} provides an overview of DESIGNATE, which consists of the following components:

\begin{figure*}[t]
    \centering
    \includegraphics[width=\textwidth]{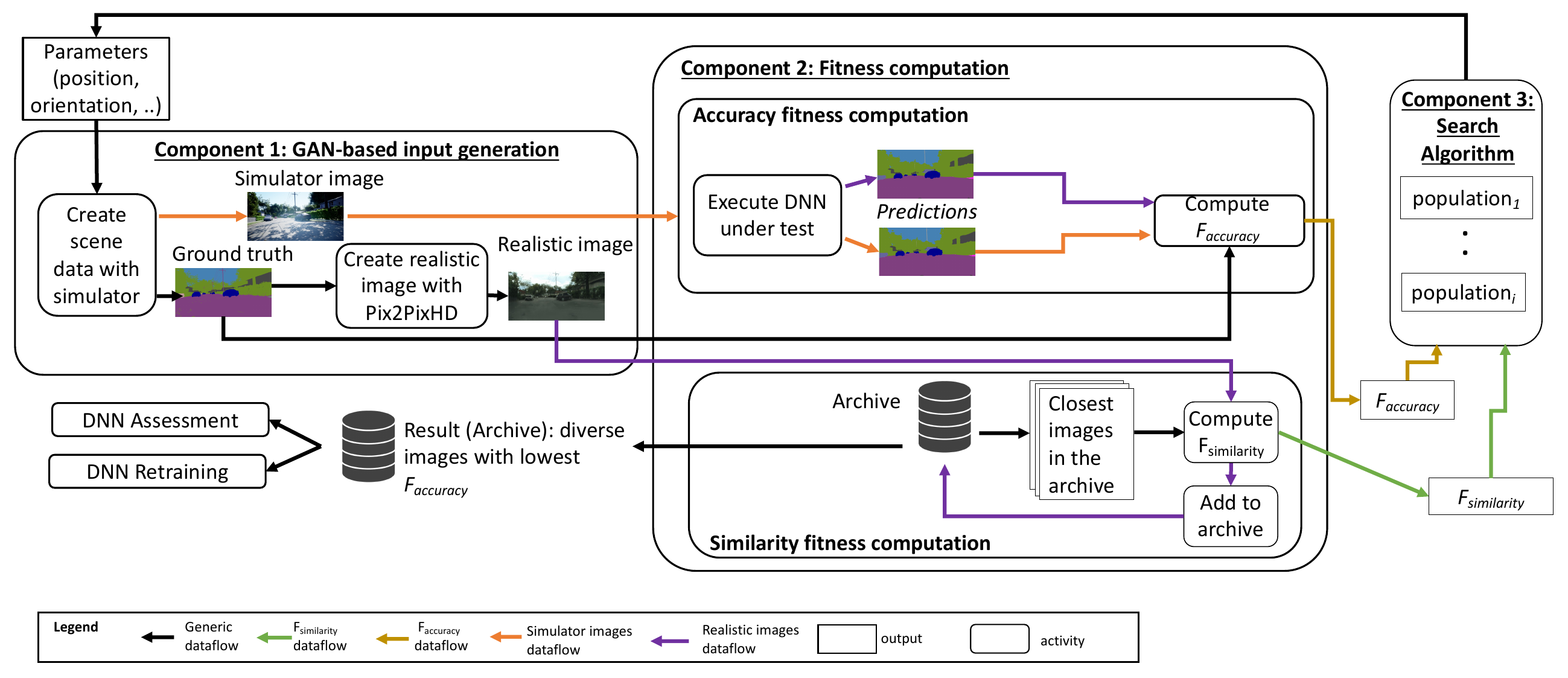}
    \caption{Overview of DESIGNATE.}
    \label{fig:designate}
\end{figure*}

\subsubsection{GAN-based Input Generation} 
\begin{figure}[h]
\includegraphics[width=8.8cm]{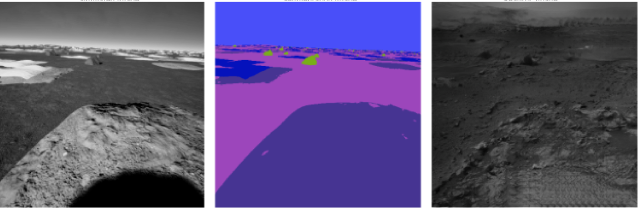}
    \caption{Examples of a Mars simulator's simulated image, its ground truth, and the realistic image generated from it by Pix2PixHD.}
    \label{fig:mars:gan:example}
\end{figure}

it leverages a GAN to generate realistic images from the simulator output. In our implementation of \DESIGNATE, 
we rely on the Pix2PixHD GAN to translate the segmentation maps (i.e., the ground truth) generated by the simulator into realistic images. Pix2PixHD needs to be trained with real-world data; for our experiments, we considered real-world images and corresponding segmentation maps. Figure~\ref{fig:mars:gan:example} shows an example realistic image generated by leveraging the ground truth provided by the \UMARS simulator. 

\subsubsection{Fitness Computation Component}
it evaluates the quality of the generated test cases based on two fitness functions:
\begin{itemize}
    \item $F_{accuracy}$, which measures the accuracy of the DNN’s predictions. For semantic segmentation tasks, it matches the $mIoU$ (see~\ref{sec:background:segmentation}), except that for images that are detected as not relevant (e.g., the sky occupies 70\% of a Mars image), a high value is returned:
    \begin{equation}
\label{eq:f1b}
\small
F_{accuracy} (i) =  
    \begin{cases}
      2 & \text{if } \mathit{skyPixelsProportion}\ge{0.7}\\
      mIoU(i) & \text{otherwise}
    \end{cases}   
\end{equation}
    \item $F_{similarity}$, which  computes the feature-based distance between a newly generated image and those already in the archive. The distance is  calculated as the Euclidean distance in a feature space derived from a ResNet50 DNN pre-trained on ImageNet~\cite{deng2009imagenet}. Further, to prevent the search algorithm from being trapped in local optima, we leverage a predefined threshold to return a high fitness value when an image is too similar to another image in the archive. For a new input $I_i$ and an archive of existing inputs $\mathcal{A}$ with feature vectors $V(I_a)$, we have:

\begin{equation}
\label{eq:f2}
\begin{aligned}
F_{similarity} (i) =  
    \begin{cases}
      2 ~~ \text{if } \mathit{distanceFromClosest}(I_i) < T \\
    \frac{1}{1 + \mathit{distanceFromClosest}(I_i)} ~~ \text{otherwise}
    \end{cases}   
\end{aligned}
\end{equation}

where:
\begin{equation*}
\mathit{distanceFromClosest}(I_i) = \min_{I_a \in \mathcal{A}} \|V(I_i) - V(I_a)\|
\end{equation*} 
with $T$ being the threshold value computed as the average pair-wise distance in a large (1000) set of randomly-generated images.

This formulation ensures that the newly generated inputs are distinct from the archived ones, promoting exploration of diverse input spaces.
\end{itemize}
    

\subsubsection{Search Algorithm}
it consists of the NSGA-II algorithm~\cite{deb2002fast, attaoui2020multi, attaoui2023improved} extended with an archive. It operates as follows:
\begin{itemize}
    \item \emph{Initialization:} A population of individuals is generated, each individual represents a specific configuration of simulation parameters.
    \item \emph{Evaluation:} Each individual is used to generate a test image, which is then provided to the DNN under test; the DNN prediction is then provided to the fitness computation component, which assigns fitness scores to the individual.
    \item \emph{Selection:} Non-dominated sorting is applied to rank individuals based on their fitness scores. A crowding distance metric is used to maintain diversity within the population.
    \item \emph{Crossover and Mutation:} Genetic operators such as simulated binary crossover and polynomial mutation are applied to create a new generation of individuals.
    \item \emph{Archiving:} An image is added to the archive if it is sufficiently distant (i.e., above $T$) from the closest image in the archive. 
    Otherwise, the image either replaces the closest image (if it has a lower $F_{accuracy}$) or it is discarded. Our strategy leads to an archive with the best-performing diverse individuals.
    \item \emph{Termination:} the algorithm terminates when a test budget (number of iterations) is exhausted.
\end{itemize}


\subsection{Oracles for automated DNN Testing}
\label{sec:background:oracles}
DNNs are usually tested using a set of inputs for which the expected output (i.e., the ground truth) has been manually specified; in our Mars example, it is a set of Mars landscape images with a corresponding segmentation map that was manually created. However, when DNN inputs are automatically generated, manual labeling is not a feasible option (e.g., because of the many generated images that make the labeling cost excessive).
For this reason, research on automated DNN testing often utilizes simulators capable of generating the required ground truth~\cite{Gambi2019,zohdinasab2021deephyperion,Haq:2021,MORLOT}, which is not a condition in our context (i.e., we deal with simulators that can't generate the needed ground truth). However, the lack of ground truth is a problem faced by work on DNN robustness testing, DNN test input prioritization, and DNN failure detection. Below, we provide an overview of how existing works address the lack of ground truth; the interested reader can refer to a recent survey for additional details~\cite{Hu2024}.

Work on \emph{DNN robustness testing} addresses the lack of ground truth by leveraging metamorphic testing. It involves modifying DNN inputs in a manner that should not affect the DNN's predictions, and subsequently verifying that the predictions remain consistent despite the alterations. For example, DeepTest~\cite{tian2018deeptest} implements programmatic image transformations that modify attributes such as blurriness, brightness, contrast, rotation angle, scale, image shearing, and translation; in all these cases the DNN prediction is expected to remain unchanged.  DeepTest  also implements rain and fog effects, while DeepRoad~\cite{zhang2018deeproad} and TACTIC~\cite{li2021testing} leverage a GAN to introduce a broader range of weather effects (e.g., rain, fog, snow, sunshine, nightlight) that are more realistic than programmatic ones; these  weather effects shall not change DNN predictions.
Other metamorphic testing approaches leverage pixel-level changes such as adding noise, 
distortions, or changing frequency and intensity~\cite{naidu2021metamorphic,missaoui2023semantic}; other works perform horizontal and vertical flipping~\cite{MTimageICSE2020}. 

Both \emph{DNN test input prioritization} and 
\emph{DNN failure detection} leverage  metrics for DNN uncertainty estimation.
A broad overview of these metrics is provided by a recent survey~\cite{UncertaintySurvey}, which enabled us to identify four categories: internal methods, external methods, Bayesian methods, and ensemble methods. 

\emph{Internal methods} leverage information about the internal functioning of the DNNs (e.g., loss or neuron activations); some of these methods are adopted at training time (e.g., DeepDiagnosis~\cite{DeepDiagnosis}) and thus out of scope because we test  trained DNNs. Among internal methods,  surprise adequacy (SA)~\cite{KimSurprise2023} is a popular method in the software engineering community. It measures how much the activation patterns of the test input deviate from those observed in the training set; it is implemented by three metrics: 
Distance-based surprise adequacy (DSA), which measures the Euclidean distance between the neuron activations for new predictions and those observed for the closest input in the training set;
Likelihood Surprise Adequacy (LSA), which leverages gaussian kernel-density estimation to compute the log-likelyhood of the observed activations; 
Mahalanobis Distance-Based Surprise Adequacy (MDSA), which leverages the Mahalanobis distance to compute the distance between a new prediction and a set of inputs in the training set. SA metrics have been used for test input selection~\cite{kim2020reducing}.
Besides SA, a number of approaches leverage the analysis of neuron activations~\cite{Usman2023,Guo2024}; however, multiple works demonstrate their limited usefulness for the identification of failure-inducing test inputs or model retraining~\cite{Canada,CoverageLo}.


\emph{External methods} leverage additional machine learning models to predict DNN uncertainty or failures; three representative approaches developed by the software engineering community are SelfOracle~\cite{stocco2020misbehaviour}, DeepGuard~\cite{Hussain2022}, and ThirdEye~\cite{Stocco2022}.  Both SelfOracle and DeepGuard report a DNN failure when the reconstruction error of an autoencoder trained on the same dataset of the DNN under test exceeds a given threshold. ThirdEye, instead, leverages a variational autoencoder trained on prediction heatmaps.

Among \emph{Bayesian methods}, a popular choice is Monte Carlo Dropout (MCD), which consists of making multiple predictions using different dropout masks (i.e., eliminating some neurons) each time; the variance across predictions is then used as a measure of uncertainty~\cite{gal2016dropout}.

\emph{Ensemble methods} consist of training multiple models and  leverage variance across predictions to determine uncertainty~\cite{lakshminarayanan2017simple}. A recent empirical assessment of an ensemble method, an MCD application, and two external methods (i.e., SelfOracle and ThirdEye) led to the better performance of the ensemble method~\cite{Grewal2024}. However, for application scenarios similar to ours (i.e., when inaccurate predictions are due to underrepresented scenarios in the training set), the ensemble method and MCD perform similarly; further, in such context, MCD outperforms external methods\footnote{See the results obtained with mutants subjects in Table~1~\cite{Grewal2024}.}. In addition, one limitation of ensemble methods is that they require the retraining of multiple DNNs under test, which is not always feasible either because of training cost or because the DNN under test is trained by a third-party. 

%% file: approach.tex
\section{The \APPR Approach}
\label{sec:approach}

\begin{figure*}[t]
    \centering
    \includegraphics[width=\textwidth]{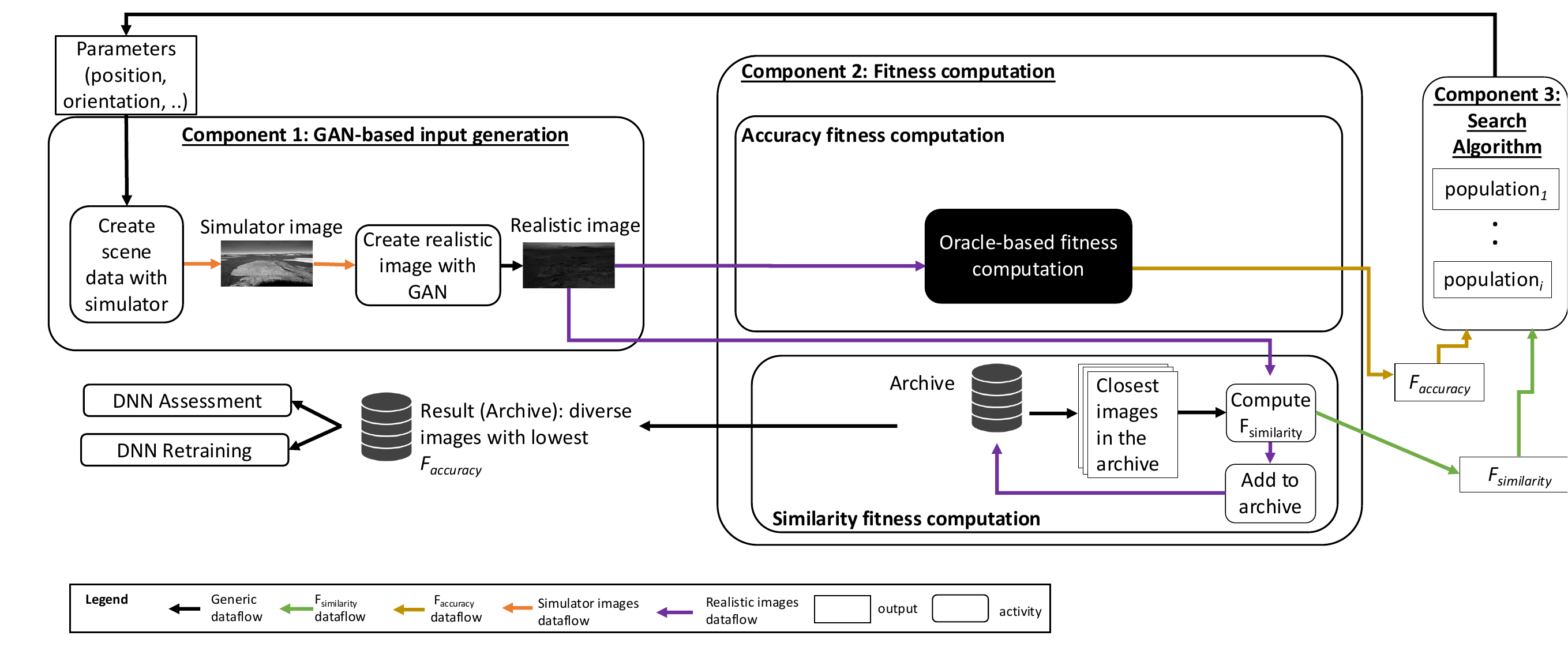}
    \caption{Overview of \APPR.}
    \label{fig:approach}
\end{figure*}

\APPR extends \DESIGNATE by replacing the reliance on ground truth with alternative oracle-based accuracy metrics for fitness computation. This enhancement enables effective DNN testing and retraining in the absence of ground-truth data.

Figure \ref{fig:approach} provides an overview of \APPR, which consists of three components: 
\emph{GAN-based Input Generation}, which receives as input a set of simulator parameter values and generates an image by leveraging a simulator whose output is enhanced by a GAN, \emph{Fitness Computation component}, which leverages DNN testing oracles (i.e., flipping, noise, SA, and MCD) to compute the fitness values required to drive the search, \emph{Search Algorithm component}, which leverages the NSGA-II multi-objective  algorithm to identify diverse and failure-inducing test cases.
\APPR's output is an archive containing a set of realistic images that likely lead to DNN failures. They can be visualized to determine the situations in which the DNN may fail~\cite{attaoui2024designate} or they can be used as training set to retrain the DNN.

The following subsections describe each component in detail along with the envisioned application scenarios.

\subsection{GAN-based Input Generation}
The GAN-based Input Generation component matches \DESIGNATE's, except for the type of GAN being used. Both \DESIGNATE and \APPR leverage a simulator to generate synthetic images based on parameterized configurations (e.g. position, orientation). These images are enhanced by a GAN to produce high-fidelity inputs. \DESIGNATE, however, uses Pix2PixHD to convert simulator-generated ground-truth data (e.g., segmentation maps) into realistic images. \APPR, which cannot leverage  ground-truth data, translates directly the simulated image into a realistic image by leveraging CycleGAN. By directly translating the ground truth data into a realistic image, Pix2PixHD provides a stronger guarantee that the generated image aligns with the provided ground truth thus reducing the risk that the search algorithm selects certain configuration parameters but they have no effect on the testing output; for example, the search may select a rover orientation that makes a small rock visible but the GAN does not show the rock. We analyze the impact of the chosen GAN on testing and retraining performance in see Section~\ref{sec:evaluation:RQ4}.

The CycleGAN model used in this step needs to be trained, in a setup phase, on a dataset consisting of simulator images (i.e., random images generated by a simulator) and real-world images (i.e., the training set of the DNN under test). 

\subsection{Fitness Computation}
\label{sec:approach:oracles}

Like \DESIGNATE, \APPR leverages two fitness functions: $F_{similarity}$, to obtain images that are diverse, and $F_{accuracy}$, to obtain images that lead to DNN failures. Since $F_{similarity}$ does not leverage ground truth information (i.e., it computes the feature-based distance between two realistic images), \APPR leverages the same $F_{similarity}$  function of \DESIGNATE (see Section~\ref{sec:background:designate}). \DESIGNATE's $F_{accuracy}$, instead, leverages ground truth information. For this reason, we defined a set of \emph{accuracy metrics} that leverage the oracle strategies proposed in the literature (see Section~\ref{sec:approach:oracles}). 
In Figure~\ref{fig:approach}, the accuracy computation is captured by the oracle-based fitness computation activity, whose returned fitness values depend on the selected accuracy metric. 

As for \DESIGNATE, the accuracy fitness shall hinder the generation of images that are not relevant for testing; 
for example, out-of-domain images. Thus, rather than directly using the computed accuracy metric, the accuracy fitness assigns a high value to images deemed irrelevant, ensuring they are demoted during the search:

\begin{equation}
F_{accuracy} (i) =  
    \begin{cases}
      2 & \text{if } i \text{ is not relevant}\\
      \mathit{accuracy\_metric}(i) & \text{otherwise}
    \end{cases}   
\end{equation}

For the context of our running example (Mars segmentation), we use the same criterion adopted in designate (i.e., proportion of sky pixels, see~Section~\ref{sec:background:designate}).

In this paper, we propose four different \emph{accuracy metrics} inspired by the oracle approaches presented in Section~\ref{sec:background:oracles}. To identify the fitness functions to use, we assessed the feasibility of all the different types of approaches in Section~\ref{sec:background:oracles} (i.e., metamorphic testing, internal methods, external methods, bayesian methods, and ensemble methods). For approaches related to metamorphic testing, we excluded the ones that may not be domain-independent such as those affecting (a) image quality (i.e., blurriness, brightness, contrast, distortion, frequency), because sometimes the quality of the captured images is expected to not change, (b) geometry (rotation angle, scale, image shearing, translation), because camera configuration may not change in the field, (c) weather conditions, because they may not be applicable in certain contexts (e.g., Mars). We thus considered the two domain-independent strategies that are often used to augment training datasets to increase DNN robustness, namely Gaussian noise and flipping. As for internal approaches, we leverage DSA, which is less computational expensive than LSA~\cite{KimSurprise2023} and is adopted by more related works than MDSA. 
For bayesian methods, we considered MCD, which is a widely adopted approach. We excluded external and ensemble methods because MCD outperforms or perform similar to them in context similar to ours (see Section~\ref{sec:background:oracles}) without requiring training of additional machine learning models, which is expensive and sometimes infeasible.


In the following, we clarify how we implemented four accuracy fitness functions leveraging the four selected accuracy metrics (flipping, noise, LSA, MCD) reported above.

\begin{figure}[h]
    \centering
    \includegraphics[width=0.5\textwidth]{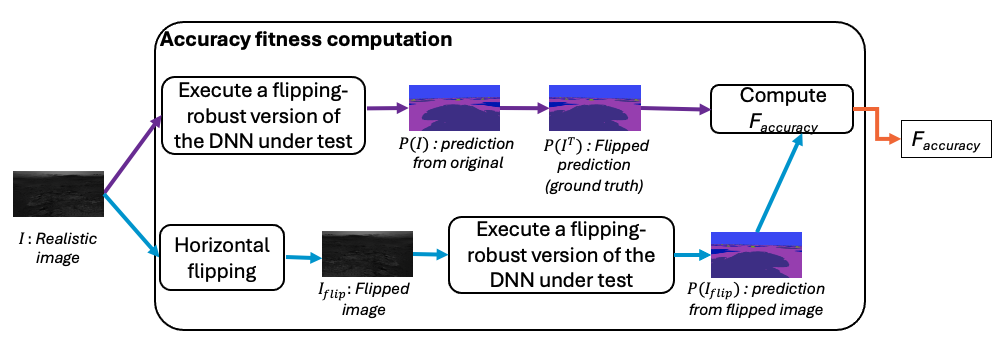}
    \caption{Fitness computation using the Flipping metric}
    \label{fig:flip-fitness}
\end{figure}

\subsubsection{Flipping Accuracy Fitness}
The flipping accuracy fitness leverages a geometric transformation to identify cases where the DNN underperforms (Figure~\ref{fig:flip-fitness}).

Given an input image $I$, it flips it horizontally to produce $I_{\text{flip}}$; then it collects the predictions (e.g., steering angle or segmentation mask) for the two images, namely,  $P(I)$ and  $P(I_{\text{flip}})$, respectively. The prediction for the original image is then transposed to produce $P(I^{T})$, which corresponds to the expected result for the flipped image and plays the role of a ground truth for accuracy computation. The transpose operation depends on the DNN under test; for example, for image segmentation DNNs it consists of flipping the image horizontally, for steering angle prediction it may consist of transposing the prediction with respect to the vertical axis. 
The accuracy fitness computes the loss between $P(I_{\text{flip}})$ and $P(I)^T$, which we use as ground truth. For segmentation DNNs, it is defined as:

\begin{equation}
F_{\text{accuracy,flip}} = \text{mIoU}(P(I)^T, P(I_{\text{flip}}))
\end{equation}

In our formula, we leverage $mIoU$ as in \DESIGNATE, where the ground truth was available. In the case of \APPR, $P(I)^T$ plays the role of the ground truth.



For an effective use of $F_{\text{accuracy,flip}}$ it is necessary to ensure that flipped images do not easily lead to DNN  failures; in our experiments, to compute $F_{\text{accuracy,flip}}$. 
instead of leveraging the original DNN under test, we used a version of the DNN under test that was retrained for 20 epochs with 50\% flipped images and 50\% non-flipped images. Note that if the DNN under test is already trained with flipped images, which is often the case to ensure robustness, such additional retraining is not needed; in our experiments, it was needed because we leveraged the same DNN used in previous work~\cite{attaoui2024designate}.

\begin{figure}[]
    \centering
    \includegraphics[width=0.5\textwidth]{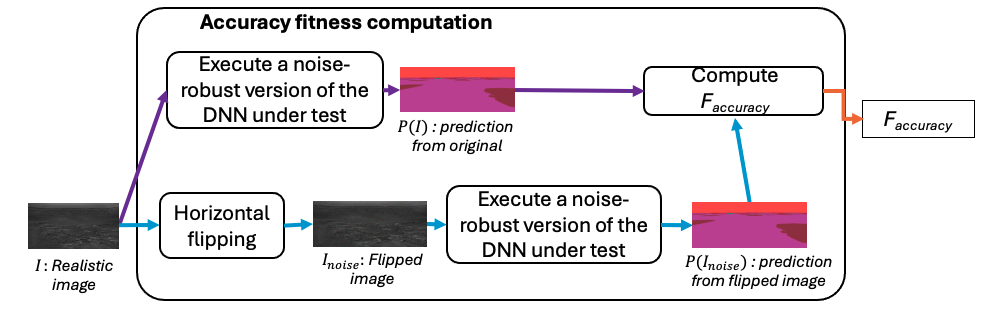}
    \caption{Fitness computation using the Noise metric}
    \label{fig:noise-fitness}
\end{figure}

\subsubsection{Noise Accuracy Fitness}
The noise accuracy fitness aims at identifying conditions where the DNN underperforms under slight perturbations, specifically, Gaussian noise \cite{zhou2019metamorphic}. 

Gaussian noise $N \sim \mathcal{N}(0, \sigma^2)$ is added to the input image $I$, resulting in a perturbed image $I_{\text{noise}}$. The predictions for the original and noisy images, $P(I)$ and $P(I_{\text{noise}})$ respectively, are compared to compute the accuracy fitness (see Figure~\ref{fig:noise-fitness}). For segmentation DNNs, the fitness can be implemented as follows:
\begin{equation}
F_{\text{accuracy,noise}} = \text{mIoU}(P(I), P(I_{\text{noise}}))
\end{equation}
The use of $mIoU$ instead of other functions (e.g., pixel-by-pixel difference) to compute the fitness value enables normalizing the deviations for each class of interest; for example, it enables reducing the impact of deviations affecting elements that cover most of the images (e.g., soil in Mars landscape or road in city environments) but are less safety-critical than mistakes affecting other items (e.g., rocks for Mars or cars for city environments).  

For regression DNNs (e.g., steering angle prediction), the fitness can be computed as the inverse of the absolute difference between the two predictions:
\begin{equation}
F_{\text{accuracy,noise}} = \frac{1}{|P(I)-P(I_{\text{noise}})|+1}
\end{equation}


To leverage the noise accuracy fitness, similar to the case of flipping, it is necessary a DNN capable of generating correct results with noisy images.
In our experiments, for example, we improved the DNN under test by retraining it for 20 epochs  using a balanced dataset comprising 50\% noisy images and 50\% non-noisy images.

\begin{figure}[h]
    \centering
    \includegraphics[width=0.5\textwidth]{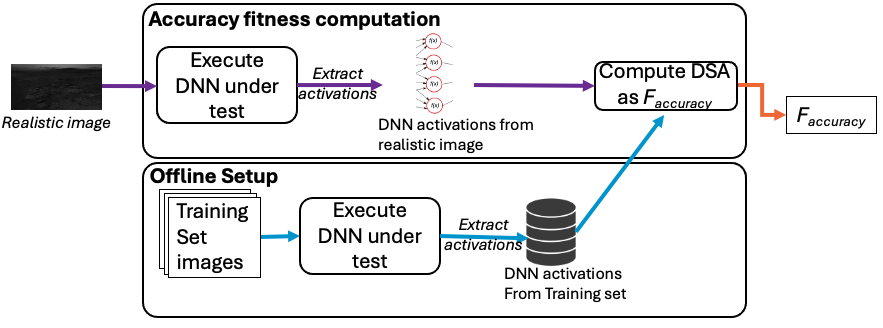}
    \caption{Fitness computation using the DSA metric}
    \label{fig:SA-fitness}
\end{figure}

\subsubsection{Surprise Adequacy Accuracy Fitness}
The Surprise Adequacy (SA) fitness shall help detecting inputs that deviate significantly from the training set’s activation patterns. The goal is to pinpoint cases where the DNN behaves unpredictably due to insufficient representation of such scenarios in the training data.

For a test input $I$, the DNN activations vector is $A(I)$. As shown in Figure~\ref{fig:SA-fitness}, this vector is compared to activation patterns in the training set, using the DSA metric, which acts as our fitness function:


\begin{equation}
F_{\text{accuracy,SA}} = \text{DSA}(I)
\end{equation}


\begin{figure}[]
    \centering
    \includegraphics[width=0.5\textwidth]{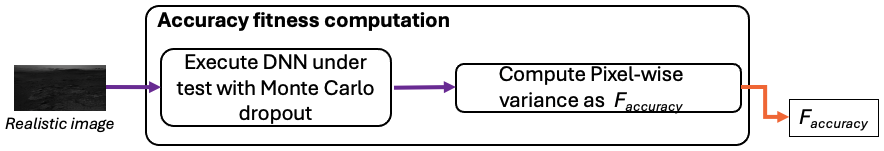}
    \caption{Fitness computation using the MCD metric}
    \label{fig:uncertainty-fitness}
\end{figure}

\subsubsection{MCD Accuracy Fitness}
The MCD fitness is designed to detect inputs where the DNN predictions are highly variable, indicating a lack of confidence. 

As shown in Figure~\ref{fig:uncertainty-fitness}, MCD is applied during $M$ stochastic forward passes of the same input $I$, generating predictions $P_1(I), P_2(I), \dots, P_T(I)$. The variability in predictions is quantified using pixel-wise variance:
\begin{equation}
\text{Var}_{\text{pixel}}(p) = \frac{1}{M} \sum_{t=1}^{M} (P_t(p) - \bar{P}(p))^2
\end{equation}
where $\bar{P}(p)$ is the mean prediction value for pixel $p$ across all passes. The MCD accuracy fitness $F_{\text{accuracy,MCD}}$ is computed as the overall uncertainty according to MCD formula:
\begin{equation}
F_{\text{accuracy,MCD}} = \frac{1}{|I|} \sum_{p \in I} \text{Var}_{\text{pixel}}(p)
\end{equation}

\subsection{Search Algorithm}

Like DESIGNATE, \APPR leverages NSGA-II extended with an archive (see Section~\ref{sec:background:oracles}).
\APPR's output corresponds to the content of the archive when NSGA-II terminates. The archive is expected to contain diverse, realistic, failure-inducing images. 

\subsection{\APPR usage scenarios}
\APPR is expected to be used with simulators that cannot generate ground-truth data, which implies that \APPR's output images do not have an associated ground truth and, consequently, can't be directly used for DNN testing or retraining. To address this situation, we envision two usage scenarios that we refer as \emph{DNN safety analysis} and \emph{DNN improvement}; they are described below. 

\emph{DNN safety analysis.} If engineers are interested in performing safety analysis, specifically, determining what are the scenarios in which the DNN may fail (e.g., to decide if they may lead to hazard or identify countermeasures), 
we recommend identify groups of similar images by leveraging clustering-based analysis pipelines~\cite{SAFE,attaoui2024supporting}. Engineers can then inspect a few (e.g., five) images in each cluster, manually determine their ground truth, and determine the prediction error. Such analysis shall enable engineers to decide if, for each  scenarios in which the DNN may fail, the prediction errors are acceptable; for example, because the chances of observing the scenario are low or because the prediction errors in those scenarios cannot lead to any hazard (e.g., the DNN erroneously predicts the presence of a rock when there is only sand in the landscape but this mistake simply leads to a slight path deviation).

\emph{DNN improvement.} The images generated by \APPR can be used to retrain the DNN but ground-truth data would be needed. For this reason, we suggest engineers to manually determine the ground truth for all the images in the archive before retraining. Although this process might be expensive, it is more cost-effective than labeling real-world images because (1) it does not require the collection of images from the field (e.g., in space) and (2) the search-based exploration in \APPR helps selecting images that are likely failure-inducing, while field collection will simply lead to randomly selected images. Further, ground-truth data generation might be simplified thanks to the use of ad-hoc tools (e.g., for image segmentation) and crowdsourcing~\cite{Crowdoracles,riccio2019deepjanus}.


%% file: evaluation.tex
\section{Empirical Evaluation}
\label{sec:evaluation}
We evaluate the effectiveness of \APPR in a Martian environment and address the following research questions (RQs):

\subsection{Research Questions}
\begin{itemize}
    \item[\textbf{RQ1}] \textbf{How do \APPR's accuracy fitnesses compare to accuracy fitnesses based on ground truth, for DNN testing?}
    Fitness functions not driven by ground truth may lead to the generation of images that may optimize the selected fitness function (e.g., different predictions when flipped) but do not help spotting limitations that are detected with a fitness based on ground truth (i.e., images leading to bad predictions). We aim to identify the best fitness for our testing purposes and compare them with a fitness based on ground truth (i.e., DESIGNATE). 

    \item[\textbf{RQ2}] \textbf{How do \APPR's accuracy fitnesses compare to accuracy fitnesses based on ground truth for DNN retraining?}
    We aim to identify the \APPR fitness leading to an archive of images that improve the DNN the most, when used for retraining, and compare with DESIGNATE's.

    \item[\textbf{RQ3}] \textbf{How does the search-based generation of realistic images without ground truth compare to the search-based generation of realistic images with ground truth for DNN testing and retraining?} Without the possibility to leverage ground truth to generate realistic images, GANs may generate images that differ from what intended by the simulator thus leading to ineffective testing; for example, the search may lead to a scenario with small rocks on the far background of a sand landscape but the GAN generates realistic images where small rocks are not visible and thus does not help discriminating between cases where rocks are correctly identified or not. We therefore compare the results obtained by GANs that leverage (i.e., Pix2PixHD) and not leverage (i.e., CycleGAN) the ground truth to study the impact of the lack of ground-truth data on \APPR.
    
\end{itemize}
\subsection{Subjects of the Study}
The subject of this study is a DeepLabV3 model~\cite{chen2017deeplab}, a state-of-the-art semantic segmentation DNN, trained for the detection of soil, rocks, sand, and bedrock on Martian terrain. We trained the DNN using the AI4MARS dataset~\cite{swan2021ai4mars} for 100 epochs. We used 16064 images for training and 966 images for testing. 
In our test set, the subject DNN achieved an average \mIoU of 0.49 (min = 0.0, first-quartile = 0.1
, third-quartile = 0.77, max = 0.99).
The subject DNN was developed as part of a collaborative project with ESA.

To apply \APPR and \DESIGNATE, we leverage \MARSSIM (see Section~\ref{sec:background:simulator}).



\subsection{Assessed techniques}

To address RQ1 and RQ2, we compare the performance of \APPR variants with DESIGNATE and a  \emph{Random} baseline that  generates realistic images by randomly selecting \MARSSIM parameters and processing its outputs with the GAN. Note that our previous work~\cite{attaoui2024designate} has demonstrated that \DESIGNATE outperforms other GAN-based testing approaches (i.e., TACTIC~\cite{zhang2018tactic}) and approaches integrating different search algorithms (i.e., DeepJanus~\cite{riccio2019deepjanus}), even when these are extended with GANs (i.e., DeepJanus-GAN); therefore, we do not compare \APPR with those approaches.

Each \APPR variant leverages $F_{\text{similarity}}$ to maximize diversity and integrates a specific $F_{\text{accuracy}}$ function (see Section~\ref{sec:approach:oracles}). We name them according to the applied fitness function: \APPRFLIP, \APPRNOISE, \APPRSURPRISE, and \APPRUNCERTAIN.

To compare the test effectiveness of \APPR fitnesses with DESIGNATE's fitness without bias, it is necessary to employ the same GAN of \DESIGNATE (i.e., Pix2PixHD). However,  Pix2PixHD requires ground-truth data to generate images; for this reason, when evaluating RQ1 and RQ2, we still leverage ground-truth data to generate realistic images, although \APPR does not leverage the ground truth to drive the search. 

For RQ3, we eliminate the reliance on ground truth by leveraging CycleGAN to transform simulator images into realistic images. To enable the use of CycleGAN in our context, we trained it with 15453 images generated by \emph{MarSim} and the full AI4MARS training set (16064 images). 


Table~\ref{tab:parameters} provides the parameter settings for the search process in \APPR and \DESIGNATE.
Two additional parameters are needed for two specific \APPR metrics. For \APPRNOISE, we set $\sigma^2$ to $0.1$, which provides sufficient perturbation without overly distorting the input data~\cite{li2019improving}. For \APPRUNCERTAIN, we set the number of MCD iterations to $5$, which provides a reliable uncertainty estimate without significantly increasing computational overhead~\cite{gal2016theoretically}.

\input{tables/parameters}




\subsection{RQ1: Test Effectiveness}
\label{sec:evaluation:RQ1}

\subsubsection{Design and measurements}
Testing approaches for vision DNNs are effective if they generate images that are diverse and lead to inaccurate DNN predictions. We therefore compare \APPR variants with \DESIGNATE and Random in terms of accuracy (i.e., \mIoU) and diversity. 
Note that for \APPRNOISE and \APPRFLIP, the archive contains the GAN-enhanced simulator images, not the results of flipping and noise; in practice, any better performance obtained by those approaches is due to the effects of the accuracy metric on the search-driven generation process not to the alteration of the images.
We evaluate diversity across the generated images by relying on feature-based distance. Specifically, we compute the feature-based distance between every image in the archive and its closest neighbor and compare their distributions. An approach with higher diversity values is preferable as it is more effective in identifying distinct situations in which segmentations are erroneous.

To address the non-deterministic nature of all considered approaches, each approach is executed ten times. 
We discuss the significance of the differences by employing the non-parametric Mann-Whitney U-test. The U statistic is computed considering all the datapoints generated across the ten executions. Furthermore, we assess effect size using Vargha and Delaney's $\hat{A}_{12}$ metric~\cite{VDA}. The $\hat{A}_{12}$ statistic, given observations (e.g., \mIoU) from two techniques X and Y, indicates the probability that technique X yields higher values than technique Y. 
When $\hat{A}_{12} < 0.50$, it indicates that X tends leading to lower values than Y, which is desirable when assessing \mIoU for testing. When $\hat{A}_{12} > 0.50$, X tends to yield higher values than Y, which is desirable when assessing feature-based distance for diversity (and \mIoU for retraining in RQ2).
The effect size is interpreted as follows:
\begin{itemize}
    \item \textbf{Small:} $0.36 < \hat{A}_{12} \leq 0.44$ or $0.56 \leq \hat{A}_{12} < 0.64$
    \item \textbf{Medium:} $0.29 < \hat{A}_{12} \leq 0.36$ or $0.64 \leq \hat{A}_{12} < 0.71$
    \item \textbf{Large:}  $\hat{A}_{12} \leq 0.29$ or $\hat{A}_{12} \geq 0.71$
\end{itemize}

\input{tables/RQ1_accuracy}
\input{tables/RQ1_accuracy_stats}

\subsubsection{Results for Accuracy}

Table~\ref{tab:rq1_quality} shows the \mIoU values obtained for each approach. Lower \mIoU values indicate better capability in revealing DNN accuracy issues. Table~\ref{tab:rq1_quality_stats} presents the statistical comparisons.

Among \APPR variants, the best one is {\APPRFLIP} which leads to a distribution closer to zero (i.e., the images generated by \APPRFLIP tend leading to worse predictions).  
{\APPRNOISE} and {\APPRSURPRISE} perform similarly, based on effect size (i.e., $\hat{A}_{12} = 0.47$). The difference with \APPRFLIP is limited (i.e., $\hat{A}_{12} = 0.55$ and $\hat{A}_{12} = 0.57$); however, \APPRFLIP clearly leads to worse predictions (i.e., all percentiles are lower). 
{\APPRUNCERTAIN} is the least effective variant, with a tangible (i.e., $\hat{A}_{12} >= 0.62$) difference from the others. We hypothesize that 
\APPRFLIP is better than \APPRUNCERTAIN because its fitness measures a tangible problem (i.e., the flipped predictions do not match and thus one of the two is wrong), while there is no guarantee that a low \APPRUNCERTAIN leads to an erronous prediction. 
\APPRNOISE is better than \APPRUNCERTAIN because noise may accentuate the mistakes of the DNN that might  be present (but with a lesser extent) in the images without noise; \APPRSURPRISE is based on distance from dataset images, which could thus be better at driving the simulator towards the generation of images that differ from the ones in the training set and where the DNN may make erroneous predictions.

All \APPR variants perform significantly better than Random, thus justifying their adoption. 

Except for \APPRUNCERTAIN, if we look at effect size, all \APPR variants do not differ from \DESIGNATE (i.e., $0.45 \le \hat{A}_{12} \le 0.52$), which suggests that three of the proposed metrics are an effective solution when ground-truth data is not available. 

\subsubsection{Results for Diversity}
\input{tables/RQ1_diversity}
\input{tables/RQ1_diversity_stats}

Table~\ref{tab:rq1_diversity} provides descriptive statistics for the diversity metrics across the generated images, while Table~\ref{tab:rq1_diversity_stats_mars} reports the corresponding $\hat{A}_{12}$  and p-values for statistical comparisons.

All \APPR variants perform better than random (higher median and avg. diversity, significant differences for p-value and $\hat{A}_{12}$), which shows that they do not undermine the search effectiveness (i.e., they still enable the search algorithm to search for diverse inputs). 

Looking at p-values and $\hat{A}_{12}$, all the \APPR variants lead to higher diversity than  \DESIGNATE ($\hat{A}_{12} \ge 0.57$), with \APPRSURPRISE outperforming the others ($0.60 \le \hat{A}_{12} \le 0.62$). The better performance of \APPRSURPRISE might be due to  $F_{accuracy,SA}$ help retaining all the inputs that are different (neuron activations) from the ones in the training set, although the retained inputs may not necessarily spot critical performance issues (see RQ1), thus resulting in slightly higher diversity. 


\begin{tcolorbox}[boxsep=0mm, left=1mm, right=1mm, top=1mm, bottom=1mm]
\textbf{RQ1.} \APPRFLIP, \APPRSURPRISE and \APPRNOISE are appropriate choices for DNN testing because they achieve similar ($\hat{A}_{12}$) accuracy and diversity than \DESIGNATE's fitness based on ground truth. \APPRFLIP performs better in finding inputs that minimize DNN accuracy, \APPRSURPRISE maximizes input diversity. 
\end{tcolorbox}

\input{tables/rq2_results}
\input{tables/rq2_stats}

\subsection{RQ2: Retraining Effectiveness}
\label{sec:evaluation:RQ3}

\subsubsection{Design and measurements}
We aim to assess the improvement in DNN performance obtained after retraining the subject DNN for additional epochs with the images generated by the approaches considered for RQ1. For each approach, we retrain (100 epochs) the DNN with the AI4MARS training set augmented with the archive of images generated in one test run. This process results in ten retrained models per approach. To assess DNN performance improvements we rely on 
the \mIoU computed on the AI4Mars test set. 

To exclude that improvements are due to additional retraining epochs, we compare the performance of the approaches considered for RQ1 with the performance of the original DNN retrained (100 epochs) with the AI4MARS training set only.

To assess significance of the differences, since all the approaches are tested with the same set of images, we leverage the Wilcoxon signed-rank test, which is a non-parametric paired test~\cite{AB14}. For effect size, following related work~\cite{AIM}, we leverage the positive-rank sum $R^+$ and the negative-rank sum $R^{-}$; specifically, we compute $\hat{E} = \frac{R^+}{R^+ + R^-}$, so that $\hat{E}$ ranges from $0$ to $1$, with the same interpretation of $\hat{A}_{12}$.




\subsubsection{Results}
Tables~\ref{tab:rq2_results} and~\ref{tab:rq2_stats} provide descriptive statistics for $mIoU$ and statistical comparisons, respectively, for our subjects. 

All \APPR variants perform similarly ($0.5 \le \hat{E} \le 0.53$). Compared to the original DNN, three of them (\APPRNOISE, \APPRFLIP, and \APPRSURPRISE) improve the median \mIoU by 0.06, while \APPRUNCERTAIN improves the median \mIoU by 0.05; consistent with RQ1, \APPRUNCERTAIN shows slightly worse performance. 

All \APPR variants outperform retraining with AI4MARS only, achieving a median  \mIoU improvement of 0.03 (0.02 for \APPRUNCERTAIN), thus highlighting the usefulness of \APPR.

Only \APPRNOISE, \APPRFLIP, and \APPRSURPRISE perform better in median than Random, although only by 1 \mIoU point. 
However, all \APPR variants have a higher 5th percentile (by 9 \mIoU points, at least), which means that they all improve the worst performance cases thus improving the DNN safety; indeed, since it is more likely to experience hazards when predictions are particularly bad, a 5th percentile going from 0.11 to 0.20 may eliminate most critical failures (e.g., if these happen with $mIoU < 0.10$).

\DESIGNATE performs better than all \APPR variants, with a median \mIoU higher by 0.03; however, the effect size is small ($ 0.56 \le \hat{E} \le 0.58$) thus indicating that in the absence of ground truth, the \APPR fitnesses are a valid solution to drive the generation of images for retraining.

\begin{tcolorbox}[boxsep=0mm, left=1mm, right=1mm, top=1mm, bottom=1mm]
\textbf{RQ2.} In the absence of ground truth, \APPRFLIP, \APPRSURPRISE, and \APPRNOISE are an effective solution to generate images for DNN retraining. Their results are slightly worse than those obtained by \DESIGNATE's fitness based on ground truth, and provide better safety guarantees than a random baseline.
\end{tcolorbox}

\subsection{RQ3: Impact of absence of ground truth for image generation on DNN testing and retraining.}
\label{sec:evaluation:RQ4}

\subsubsection{Design and measurements}

We aim to assess if \APPR, when using a GAN that does not leverage ground-truth data (i.e., CycleGAN), achieves the same DNN testing and retraining effectiveness observed with a GAN that leverages ground-truth data (i.e., Pix2PixHD, used for RQ1 and RQ2).
We thus test the subject DNN with the same \APPR versions considered for RQ1 but after replacing Pix2PixHD with CycleGAN (we used the same simulator to not introduce bias in the assessment). We repeated each assessment ten times and compared the results with those obtained with Pix2PixHD, by collecting the metrics described in the following paragraphs.


To compare Pix2PixHD and CycleGAN for \emph{test effectiveness} we cannot leverage the \mIoU computed on the generated images (i.e., what done for RQ1) because there is no guarantee that the images generated by CycleGAN match the ground truth produced by \MARSSIM. To select alternative metrics for the comparison, we note that, in \APPR, two input generation techniques (here, CycleGAN and Pix2PixHD) behave similarly (1) if they enable achieving the same accuracy fitness and (2) if they enable \APPR to explore the same areas of the input space. Therefore, we leverage two metrics. First, we compare the fitness distribution obtained when integrating CycleGAN and Pix2PixHD into the different \APPR variants. Second, to discuss how the input space is explored by the different approaches, we apply a clustering pipeline to the archive images generated with CycleGAN and Pix2PixHD. Specifically, for each \APPR variant, we apply a clustering 
pipeline on the set of images resulting from the union of all the images generated with CycleGAN and Pix2PixHD in the ten runs. For each cluster, we count the number of images belonging to the CycleGAN and Pix2PixHD variant. The two input GANs enable inspecting the same areas of the input space if each cluster includes at least one image for each.

Image clustering is sensible to the colors present in the images and thus may aggregate together images that contain the same colors rather than images that contain the same items. For this reason, instead of clustering the images stored into the archive, we cluster their ground truth, which we kept generating with \emph{MarSim} despite we did not use it for input generation. As a clustering pipeline, we leverage a combination of the ResNet50 feature extraction model, the DBSCAN clustering algorithm, and the UMAP dimensionality reduction technique; we selected such pipeline among others~\cite{Mohammed:PipelinesAssessment} because, in this context, it led to better clusters, according to visual inspection and cluster quality metrics (DBI and Silhouette index). 

To select the metrics to assess \emph{test effectiveness} in terms of diversity we can leverage the same metrics of RQ1 and compare the results obtained with CycleGAN with the results obtained with Pix2PixHD and \DESIGNATE.

To assess \emph{retraining effectiveness}, for each of the ten runs of the \APPR variants, we use the archive images generated with CycleGAN to retrain the subject DNN following to the same procedure adopted for RQ2. For each \APPR variant, we then compare the DNN performance (i.e., \mIoU computed on the AI4MARS test set) obtained after retraining the DNN with images generated by CycleGAN and Pix2PixHD.





\subsubsection{Results for Test Effectiveness (fitness distribution)}

\begin{figure}[tb]
    \centering
    \includegraphics[width=0.5\textwidth]{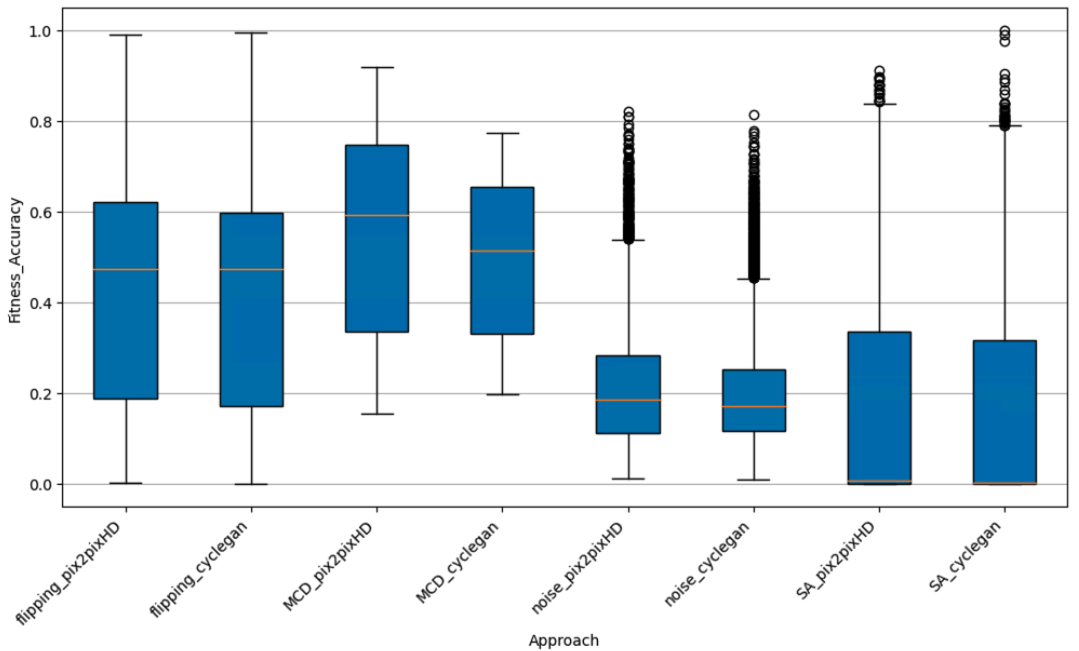}
    \caption{Comparison between each two variants of one approach (Pix2pixHD and CycleGAN) based on their respective $Fitness_{Accuracy}$ value.}
    \label{fig:rq3_fitness}
\end{figure}

Figure~\ref{fig:rq3_fitness} shows boxplots with the $F_{accuracy}$ obtained for each oracle-based approach when using Pix2PixHD and CycleGAN, respectively (each datapoint is the $F_{accuracy}$ of one image collected in one of the ten runs). Table~\ref{tab:rq3_stats} reports the corresponding $\hat{A}_{12}$ and U-test p-values for each variant pair. 

Figure~\ref{fig:rq3_fitness} shows that Pix2PixHD and CycleGAN perform similarly, which is confirmed by  Table~\ref{tab:rq3_stats} showing no tangible difference ($ 0.49 \le \hat{A}_{12} \le 0.53$) for all variants except \APPRUNCERTAIN. For \APPRUNCERTAIN,  CycleGAN leads to a lower median fitness (i.e., performs better); it could be due to CycleGAN generating images that are slightly different than the ones in the training set thus increasing DNN uncertainty.



\subsubsection{Results for Test Effectiveness (input space exploration)}

Table~\ref{tab:rq3_clusters} provides the number of clusters generated for each \APPR variant, the number and percentage of clusters covered only by one of the two approaches, and the p-value obtained with the Wilcoxon signed-rank test.

For each \APPR variant, only a few clusters contain images belonging to one GAN only. After inspecting those clusters manually, we conclude that those are clusters resulting from imprecise clustering: they are similar to images in clusters containing  images generated with both the two GANs. We can thus conclude that CycleGAN and Pix2PixHD enable inspecting the same areas of the input space. 

The Wilcoxon signed-rank test applied to the distribution of images per cluster, for \APPRFLIP and \APPRUNCERTAIN enables us to reject (p-value $< 0.01$) the null hypothesis ``the distribution of images across clusters is the same'', which suggests that, despite the two GANs still enable the fitnesses to drive the inspection of the same sub-spaces, they generate a significantly different number of images for each sub-space (i.e., explore the input space differently). For \APPRNOISE and \APPRSURPRISE, we have no evidence of a difference in space exploration between CycleGAN and Pix2PixHD (p-value $> 0.1$), suggesting that these two fitnesses lead to consistent results when used with CycleGAN and Pix2PixHD. However, one way to assess the practical consequences of such observation is by comparing how the images generated by the different approaches affect retraining, which is what we discuss in Section~\ref{sec:RQ3:retrain}\footnote{Since the results in Section~\ref{sec:RQ3:retrain} show no difference in retraining effectiveness between \APPRNOISE and \APPRFLIP, we conclude that the differences in the distribution of images across clusters have limited impact on \APPR's outcome.}.

\input{tables/rq3_pix2pix_cycleGAN_clusters}

\input{tables/rq3_stats}


    
    
    

\subsubsection{Results for Test Effectiveness (diversity)}

\input{tables/rq3_cyclegan_diversity}
\input{tables/rq3_cyclegan_diversity_stats}

Table~\ref{tab:rq3_cyclegan_diversity} and~\ref{tab:rq3_cyclegan_diversity_mars} provide descriptive statistics for test input diversity, $\hat{A}_{12}$ and U-test p-values.

\APPRSURPRISECYCLE and \APPRFLIP perform similarly; \APPRSURPRISECYCLE has a higher median (12.84), \APPRFLIP has a higher average (13.15). However, they do not differ significantly ($\mathit{p-value} > 0.05$). Further, all \APPR metrics do not differ significantly when combined with CycleGAN ($46 \le \hat{A}_{12} \le 54$).

Except for \APPRSURPRISECYCLE, using CycleGAN reduces the ability of the selected fitnesses in driving test diversity (see $\hat{A}_{12}$ values comparing $ORBIT_{\mathit{fitness},\mathit{cycleGAN}}$ and $ ORBIT_{\mathit{fitness}}$). However, this does not cause worse performances, compared to \DESIGNATE. Indeed, all the \APPR fitnesses lead to higher diversity than \DESIGNATE, with \APPRFLIP and \APPRSURPRISE having a performance that is tangibly better than \DESIGNATE's (i.e., $\hat{A}_{12} \ge 0.58$).

\subsubsection{Results for DNN retraining}
\label{sec:RQ3:retrain}

\input{tables/rq3_retraining_cyclegan}
\input{tables/rq3_retraining_cyclegan_stats}

Table~\ref{tab:rq3_cyclegan_retraining} shows the retraining performance obtained with the subject DNN retrained using the four \APPR variants relying on CycleGAN; Table~
\ref{tab:rq3_cyclegan_retraining_stats} provides $\hat{E}$ and p-values. 

The performance obtained by the \APPR variants with CycleGAN is similar to the one observed with Pix2PixHD (median is higher by 0.01 in most cases, but difference is not practically significant based on $\hat{E}$), which suggests that relying on a GAN that does not leverage the ground truth does not affect the retraining capabilities of \APPR. The best results are observed with \APPRFLIPCYCLE and \APPRNOISECYCLE, with a median and average \mIoU that are lower than \DESIGNATE only by 0.02 and 0.01, respectively. Based on Table~\ref{tab:rq3_cyclegan_retraining_stats},  there is no tangible difference between \APPRFLIPCYCLE, \APPRNOISECYCLE, \APPRUNCERTAINCYCLE, and \DESIGNATE ($\hat{E}=0.45$). 

\begin{tcolorbox}[boxsep=0mm, left=1mm, right=1mm, top=1mm, bottom=1mm]
\textbf{RQ3.} The use of a GAN that does not leverage the ground truth (e.g., CycleGAN) does not harm the effectiveness of \APPR. Indeed, the Pix2PixHD and the CycleGAN version of each \APPR fitness perform similarly. In addition, they drive the inspection of the same areas of the input space. Furthermore, with CycleGAN, all \APPR approaches perform similar to or better than \DESIGNATE in terms of test diversity. Last, CycleGAN leads to better retraining results than Pix2PixHD, with \APPRFLIPCYCLE and \APPRNOISECYCLE having the highest median and leading to an $mIoU$ similar to \DESIGNATE's.
\end{tcolorbox}

\subsection{Discussion}

RQ1 and RQ2 results show that \APPRFLIP, and \APPRSURPRISE are the best choice in absence of ground truth. Indeed, for testing, \APPRFLIP achieves lower \mIoU values than 
\DESIGNATE and other \APPR fitnesses but \APPRSURPRISE performs better than others in terms of diversity. They both perform similar to \DESIGNATE for retraining.

RQ3 results show that the \APPR performance observed for RQ1 and RQ2 is not worsened by a GAN that does not leverage ground truth. \APPRFLIP and \APPRSURPRISE remain the best performing fitnesses for testing (i.e., they achieve the highest diversity). However, the two fitnesses with the highest median and performing similar to \DESIGNATE for retraining are \APPRFLIP and \APPRNOISE, which makes \APPRFLIP the best fitness for both testing and retraining tasks. Future work will assess the combination of the different fitness functions with a many-objective search approach.

\subsection{Threats to validity}

To address threats to \emph{internal validity}, we used the implementation of the state-of-the-art subject DNN and GANs provided by their authors; further, we carefully verified our implementation for \APPR.

For \emph{construct validity}, we addressed RQ1 and RQ2 in terms of \mIoU, which is a metric for the assessment of segmentation DNNs. For RQ1 we assessed also test diversity, which we measured in terms of feature-based distance, which, in our previous work, has shown to accurately capture differences across input images~\cite{attaoui2024designate}. For RQ3, we leveraged fitness and \mIoU, which are two direct metrics to compare the performance of DNN testing approaches. To assess input space exploration we leveraged clusters, which have been adopted by related work for similar purposes~\cite{Zohdinasab2023b}.

As for \emph{generalizability}, we considered a DNN task (i.e., image segmentation) that is integrated into many autonomous systems. Further, our previous results have shown that \DESIGNATE results generalize to autonomous driving.

\subsection{Data availability}

Our replication package is available online~\cite{replicability}.

%% file: tables/parameters.tex
\begin{table}[t]
\centering
\caption{Search settings for \APPR and \DESIGNATE}
\label{tab:parameters}
\begin{tabular}{|l|l|c|}
\hline
\multicolumn{2}{|l|}{\textbf{Parameters}}              & \textbf{Value} \\
\hline
\multirow{7}{*}{Search process}&Population size & 12 \\
&Generations & 100 \\
&Mutation probability & 0.3 \\
&Crossover probability & 0.7 \\
&$T_{similarity}$ &  12 \\
\hline
\end{tabular}
\end{table}

%% file: tables/RQ1_accuracy.tex
\begin{table}[t]
\smaller
\centering
\caption{RQ1-Accuracy assessment. Descriptive statistics for $mIoU$ obtained with the generated images.}
\label{tab:rq1_quality}
\begin{tabular}{|@{}l@{}|c@{\hspace{0.5mm}}|c@{\hspace{0.5mm}}|c@{\hspace{0.5mm}}|c@{\hspace{0.5mm}}|c@{\hspace{0.5mm}}|c@{\hspace{0.5mm}}|c@{\hspace{0.5mm}}|@{}r@{\hspace{0.5mm}}|}
\hline
\textbf{} & \textbf{\#images} & \multicolumn{5}{c|}{$\mathbf{mIoU}\_\textbf{distribution}$}  \\ \hline
\textbf{} & \textbf{} &  {\textbf{median}} & {\textbf{5th perc.}} & {\textbf{1st quart.}} &  {\textbf{3rd quart.}} & {\textbf{avg.}} \\ \hline

\textbf{\APPRFLIP} & 740  & \textbf{0.44} & \textbf{0.08} & \textbf{0.29} & \textbf{0.63} & \textbf{0.46} \\ \hline

\textbf{\APPRSURPRISE} & 649  &  {0.50} &  {0.15} &  {0.36} &  {0.70} & 0.51 \\ \hline

\textbf{\APPRNOISE} & 865   &  {0.48} &  {0.16} &  {0.37} &  {0.65} & 0.50 \\ \hline

\textbf{\APPRUNCERTAIN} & 791 &    {0.63} &  {0.25} &  {0.49} &  {0.74} & 0.60 \\ \hline


\textbf{DESIGNATE} & 764  &  {0.50} &  {0.09} &  {0.33} &  {0.69} & 0.50 \\ \hline



\textbf{Random} & 1031 &   {0.64} &  {0.24} &  {0.44} &  {0.83} & 0.62 \\ \hline
\end{tabular}%
 
\vspace{1mm}
Note: best (i.e., lowest) results (per column) in bold. p. indicates percentile.
\end{table}

%% file: tables/RQ1_accuracy_stats.tex
\begin{table}[tb]
\smaller
\centering
\caption{RQ1. $\hat{A}_{12}$ for the data in Table~\ref{tab:rq1_quality}.}
\label{tab:rq1_quality_stats}
\begin{tabular}{|l|l|l|l|l|l|l|}
\hline
 & {\textbf{\FLIP}} & {\textbf{\SURPRISE}} & {\textbf{\NOISE}} & {\textbf{\UNCERTAIN}} & 
 {\textbf{DESIGN.}} & 
 {\textbf{Random}} \\
\hline
\textbf{\APPRFLIP} & --  & \textbf{0.43}  & 0.45  & \textbf{0.31} & 0.45  & \textbf{0.31}  \\
\hline
\textbf{\APPRSURPRISE} & \textbf{0.57}  & --  & 0.53  & \textbf{0.38}  & 0.52  & \textbf{0.37}  \\
\hline
\textbf{\APPRNOISE} & 0.55  & 0.47  & --  & \textbf{0.34}  & [0.49]  & \textbf{0.34}  \\
\hline
\textbf{\APPRUNCERTAIN} & \textbf{0.69}  & \textbf{0.62}  & \textbf{0.66}  & --  & \textbf{0.63}  & 0.47  \\
\hline
\textbf{DESIGNATE} & 0.55  & 0.48  & [0.51] & \textbf{0.37}  & --  & \textbf{0.36}  \\
\hline
\textbf{Random} & \textbf{0.69} & \textbf{0.63}  & \textbf{0.66}  & 0.53  & \textbf{0.64}  & --  \\
\hline

\hline
\end{tabular}
\vspace{1mm}

Note: All p-values $< 0.01$, except where `[]' is used to indicate $> 0.1$. Tangible (based on $\hat{A}_{12}$) differences in bold; for each pair of approaches, the best approach is the one with $\hat{A}_{12} < 0.50$ and its name on the row.
\end{table}

%% file: tables/RQ1_diversity.tex
\begin{table}[t]
\smaller
\centering
\caption{RQ1-Diversity assessment. Descriptive statistics for diversity across images in the archive.}
\label{tab:rq1_diversity}
\begin{tabular}{|@{}l@{}|c@{\hspace{0.5mm}}|c@{\hspace{0.5mm}}|c@{\hspace{0.5mm}}|c@{\hspace{0.5mm}}|c@{\hspace{0.5mm}}|c@{\hspace{0.5mm}}|c@{\hspace{0.5mm}}|@{}r@{\hspace{0.5mm}}|}
\hline
\textbf{} & \multicolumn{5}{c|}{\textbf{Feature Distance Diversity}} \\ \hline
 \textbf{} &  {\textbf{median}} & {\textbf{5th perc.}} & {\textbf{1st quart.}} &  {\textbf{3rd quart.}} & {\textbf{avg.}} \\ \hline

\textbf{\APPRFLIP} &  {12.94} & {12.29} & {12.61} & {13.33} & 13.20 \\ \hline

\textbf{\APPRSURPRISE} & {\textbf{13.37}} & {\textbf{12.43}} & {\textbf{12.68}} & {\textbf{14.20}} & \textbf{13.71} \\ \hline

\textbf{\APPRNOISE} & {13.01} & {12.18} & {12.47} & {13.69} & 13.28 \\ \hline

\textbf{\APPRUNCERTAIN} & {12.87} & {12.05} & {12.39} & {13.38} & 13.07 \\ \hline

\textbf{DESIGNATE} & {12.77} & {12.04} & {12.16} & {13.43} & 12.90 \\ \hline

\textbf{Random} & {5.88} & {4.66} & {5.35} & {6.53} & 6.05 \\ \hline

\end{tabular}%

Note: best (i.e., highest or most diverse) results (per column) in bold.
\end{table}










%% file: tables/rq1_diversity_stats.tex
\begin{table}[t]
\smaller
\centering
\caption{RQ1. $\hat{A}_{12}$ for the data in Table~\ref{tab:rq1_diversity}.}
\label{tab:rq1_diversity_stats_mars}
\begin{tabular}{|l|l|l|l|l|l|l|}
\hline
 {\textbf{\FLIP}} & {\textbf{\SURPRISE}} & {\textbf{\NOISE}} & {\textbf{\UNCERTAIN}} & {\textbf{DESIGN.}} & {\textbf{Random}} \\
\hline
\textbf{\APPRFLIP} & --  & \textbf{0.38}  & (0.51)  & \textbf{0.57}  & \textbf{0.60}  & \textbf{1.00}  \\ \hline
\textbf{\APPRSURPRISE} & \textbf{0.62}  & --  & \textbf{0.60}  & \textbf{0.64}  & \textbf{0.70}  & \textbf{1.00}  \\ \hline
\textbf{\APPRNOISE} & (0.49)  & \textbf{0.40}  & --  & 0.55  & \textbf{0.61}  & \textbf{1.00}  \\ \hline
\textbf{\APPRUNCERTAIN} & \textbf{0.43}  & \textbf{0.36}
 & 0.45  & --  & \textbf{0.57}  & \textbf{1.00}  \\ \hline
\textbf{DESIGNATE} & \textbf{0.40}  & \textbf{0.30}  & \textbf{0.39}  & \textbf{0.43}  & --  & \textbf{1.00}  \\ \hline
\textbf{Random} & \textbf{0.00}  & \textbf{0.00}  & \textbf{0.00}  & \textbf{0.00}  & \textbf{0.00}  & -- \\ \hline

\hline
\end{tabular}
\vspace{1mm}
\\
Note: All p-values $< 0.01$, except where `()' is used to indicate $0.05 < p-value < 0.1$. Tangible ($\hat{A}_{12}$) differences in bold; for each pair, the best approach is the one with $\hat{A}_{12} > 0.50$ and its name on the row.
\end{table}

%% file: tables/rq2_results.tex
\begin{table*}[t]
\centering
\footnotesize
\caption{RQ2. $mIoU$ results obtained with the DeeplabV3 model retrained using the outputs of the different approaches.}
\label{tab:rq2_results}
\begin{tabular}{|@{}l|l|l|l|l|l|l|l|}
\hline
\textbf{Retraining set} & \textbf{min} & \textbf{max} & \textbf{median} & \textbf{5th percentile} & \textbf{1st quartile} & \textbf{3rd quartile} & \textbf{Average} \\ \hline

\textbf{None (Original DeeplabV3)} & 0.00 & 0.99 & 0.44 & 0.01 & 0.26 & \textbf{0.77} & 0.49 \\ \hline
\textbf{AI4Mars training (Retrained DNN)} & \textbf{0.01 (+0.01)} & 1.00 (+0.01) & 0.47 (+0.03) & 0.17 (+0.16) & 0.33 (+0.07) & 0.63 (-0.14) & 0.49 (+0.00) \\ \hline
\textbf{\APPRFLIP} & 0.00 (+0.00) & 1.00 (+0.01) & 0.50 (+0.06) & 0.20 (+0.19) & 0.36 (+0.10) & 0.67 (-0.10) & 0.52 (+0.03) \\ \hline
\textbf{\APPRSURPRISE} & 0.00 (+0.00) & 1.00 (+0.01) & 0.50 (+0.06) & 0.19 (+0.18) & 0.36 (+0.10) & 0.67 (-0.10) & 0.52 (+0.03) \\ \hline
\textbf{\APPRNOISE} & 0.00 (+0.00) & 1.00 (+0.01) & 0.50 (+0.06) & \textbf{0.21 (+0.20)} & 0.36 (+0.10) & 0.67 (-0.10) & 0.52 (+0.03) \\ \hline
\textbf{\APPRUNCERTAIN} & 0.00 (+0.00) & 1.00 (+0.01) & 0.49 (+0.05) & 0.20 (+0.19) & 0.35 (+0.09) & 0.66 (-0.11) & 0.51 (+0.02) \\ \hline
\textbf{DESIGNATE} & 0.00 (+0.00) & 1.00 (+0.01) & \textbf{0.53 (+0.09)} & 0.19 (+0.18) & \textbf{0.38 (+0.12)} & \textbf{0.69 (-0.08)} & \textbf{0.53 (+0.04)} \\ \hline
\textbf{Random} & 0.00 (+0.00) & 1.00 (+0.01) & 0.49 (+0.05) & 0.11 (+0.10) & 0.33 (+0.07) & 0.68 (-0.09) & 0.51 (+0.02) \\ \hline

\end{tabular}
\vspace{1mm}
\\
Note: The difference with the original DeeplabV3 is shown in parentheses. Best results (per column) are in bold.
\end{table*}

%% file: tables/rq2_stats.tex
\begin{table*}[t]
\centering
\smaller
\caption{RQ2. p-values and $\hat{E}$ for the data in Table~\ref{tab:rq2_results}.}
\label{tab:rq2_stats}
\resizebox{\textwidth}{!}{%
\begin{tabular}{|l|ll|ll|ll|ll|ll|ll|ll|ll|}
\hline
 & \multicolumn{2}{c|}{\textbf{None}} & \multicolumn{2}{c|}{\textbf{AI4Mars}} & \multicolumn{2}{c|}{\textbf{\APPRFLIP}} & \multicolumn{2}{c|}{\textbf{\APPRSURPRISE}} & \multicolumn{2}{c|}{\textbf{\APPRNOISE}} & \multicolumn{2}{c|}{\textbf{\APPRUNCERTAIN}} & \multicolumn{2}{c|}{\textbf{DESIGNATE}} & \multicolumn{2}{c|}{\textbf{Random}} \\
\hline
 & \textbf{$\hat{E}$} & \textbf{p-value} & \textbf{$\hat{E}$} & \textbf{p-value} & \textbf{$\hat{E}$} & \textbf{p-value} & \textbf{$\hat{E}$} & \textbf{p-value} & \textbf{$\hat{E}$} & \textbf{p-value} & \textbf{$\hat{E}$} & \textbf{p-value} & \textbf{$\hat{E}$} & \textbf{p-value} & \textbf{$\hat{E}$} & \textbf{p-value} \\
\hline
\textbf{None} & -- & -- & 0.48& 1.67E-03& \textbf{0.42}& 1.50E-40& \textbf{0.42}& 4.37E-35& \textbf{0.42}& 1.05E-45& \textbf{0.43}& 3.69E-37& \textbf{0.38}& 1.48E-99& \textbf{0.48}& 2.84E-05\\ \hline
\textbf{AI4Mars} & 0.52& 1.67E-03& -- & -- & 0.47& 1.39E-06& 0.48& 2.83E-04& 0.47& 2.02E-05& 0.49& 2.22E-02& 0.44& 7.13E-26& 0.49& 2.31E-01\\ \hline
\textbf{\APPRFLIP} & \textbf{0.58}& 1.50E-40& 0.53& 1.39E-06& -- & -- & 0.5& 7.26E-01& 0.49& 2.31E-01& 0.52& 2.98E-05& \textbf{0.44}& 3.08E-26& 0.54& 2.66E-12\\ \hline
\textbf{\APPRSURPRISE} & \textbf{0.58}& 4.37E-35& 0.52& 2.83E-04& 0.5& 7.26E-01& -- & -- & 0.5& 6.15E-01& 0.53& 1.16E-04& \textbf{0.44}& 4.17E-19& 0.55& 1.95E-12\\ \hline
\textbf{\APPRNOISE} & \textbf{0.58}& 1.05E-45& 0.53& 2.02E-05& 0.51& 2.31E-01& 0.5& 6.15E-01& -- & -- & 0.53& 1.22E-07& \textbf{0.44}& 1.64E-27& 0.54& 2.00E-11\\ \hline
\textbf{\APPRUNCERTAIN} & \textbf{0.57}& 3.69E-37& 0.51& 2.22E-02& 0.48& 2.98E-05& 0.47& 1.16E-04& 0.47& 1.22E-07& -- & -- & \textbf{0.42}& 1.27E-41& 0.52& 3.68E-05\\ \hline
\textbf{DESIGNATE} & \textbf{0.62}& 1.48E-99& \textbf{0.56}& 7.13E-26& \textbf{0.56}& 3.08E-26& \textbf{0.56}& 4.17E-19& \textbf{0.56}& 1.64E-27& \textbf{0.58}& 1.27E-41& -- & -- & \textbf{0.59}& 2.94E-49\\ \hline
\textbf{Random} & 0.52& 2.84E-05& 0.51& 2.31E-01& 0.46& 2.66E-12& 0.45& 1.95E-12& 0.46& 2.00E-11& 0.48& 3.68E-05& 0.41& 2.94E-49& -- & -- \\ \hline
\end{tabular}%
}
\vspace{1mm} \\
Note: Tangible ($\hat{E}$) differences in bold; for each pair, the best approach is the one with $\hat{E} > 0.50$ and its name on the row.
\end{table*}

%% file: tables/rq3_pix2pix_cycleGAN_clusters.tex
\begin{table}[t]
\centering
\caption{RQ3: input space exploration. Number of clusters generated for each
\APPR variants, number (percentage) of clusters that contain only images of one approach, and p-value.}
\label{tab:rq3_clusters}
\begin{tabular}{|@{}l@{\hspace{0.5mm}}|@{}r@{\hspace{0.5mm}}|@{}r@{\hspace{0.5mm}}|@{}r@{\hspace{0.5mm}}|@{}r@{\hspace{0.5mm}}|}
\hline
Approach                                           & Total Clusters & Pix2pixHD only& CycleGAN only& p-value\\ \hline
\APPRFLIP& 26             & 2 (7.69\%)                                     & 0 (0.00\%)                                    & 6.30E-03                   \\ \hline

\APPRSURPRISE & 34             & 2 (5.88\%)                                     & 1 (2.94\%)                                    & 8.51E-01                   \\ \hline

\APPRNOISE & 123            & 2 (1.63\%)                                     & 17 (13.82\%)                                   & 3.19E-01                   \\ \hline

\APPRUNCERTAIN & 31             & 0 (0.00\%)                                     & 2 (6.45\%)                                    & 1.11E-02                   \\ \hline
\end{tabular}%

\end{table}

%% file: tables/rq3_stats.tex
\begin{table}[tb]
\centering

\caption{RQ3: test effectiveness. $F_{accuracy}$: median, $A_{12}$, and p-values for the data in Figure~\ref{fig:rq3_fitness}.}
\label{tab:rq3_stats}
\resizebox{0.5\textwidth}{!}{%
\begin{tabular}{|l|l|l|c|c|}

\hline

\textbf{Variant }&\textbf{Pix2PixHD}             & \textbf{CycleGAN}              & \textbf{VDA}  & \textbf{p-value}  \\ \hline
\APPRFLIP& 0.48    & 0.47     & 0.51 & 1.67E-02 \\ \hline

\APPRSURPRISE& 0.009          & 0.004           & 0.49 & 2,58E-01 \\ \hline

\APPRNOISE& 0.18       & 0.17        & 0.53 & 5.78E-13 \\ \hline

\APPRUNCERTAIN& 0.59 & 0.51  & 0.59 & 1,09E-81 \\ \hline

\end{tabular}%
}
\end{table}

%% file: tables/rq3_cyclegan_diversity.tex
\begin{table}[t]
\smaller
\centering
\caption{RQ3-Diversity assessment. Descriptive statistics for diversity across images generated with CycleGAN.}
\label{tab:rq3_cyclegan_diversity}
\begin{tabular}{|@{}l@{}|c@{\hspace{0.5mm}}|c@{\hspace{0.5mm}}|c@{\hspace{0.5mm}}|c@{\hspace{0.5mm}}|c@{\hspace{0.5mm}}|c@{\hspace{0.5mm}}|c@{\hspace{0.5mm}}|@{}r@{\hspace{0.5mm}}|}
\hline
\textbf{} & \multicolumn{5}{c|}{\textbf{Feature Distance Diversity}} \\ \hline
 \textbf{} &  {\textbf{median}} & {\textbf{5th perc.}} & {\textbf{1st quart.}} &  {\textbf{3rd quart.}} & {\textbf{avg.}} \\ \hline

\textbf{\APPRFLIPCYCLE} & 12.75 & \textbf{12.12} & 12.40 & \textbf{13.68} & \textbf{13.15} \\ \hline
\textbf{\APPRSURPRISECYCLE} & \textbf{12.84} & 12.11 & \textbf{12.49} & 13.41 & 13.10 \\ \hline
\textbf{\APPRNOISECYCLE} & 12.67 & 12.07 & 12.26 & 13.51 & 13.08 \\ \hline
\textbf{\APPRUNCERTAINCYCLE} & 12.68 & 12.03 & 12.38 & 13.58 & 13.05 \\ \hline
\end{tabular}
Note: best (i.e., highest) results (per column) in bold.
\end{table}

%% file: tables/rq3_cyclegan_diversity_stats.tex
\begin{table}[t]
\smaller
\centering
\caption{RQ3. $A_{12}$ for the data in Table~\ref{tab:rq3_cyclegan_diversity}.}
\label{tab:rq3_cyclegan_diversity_mars}

\begin{tabular}{|@{}l@{}|@{\hspace{0.5mm}}c@{\hspace{0.5mm}}|@{\hspace{0.5mm}}c@{\hspace{0.5mm}}|c@{\hspace{0.5mm}}|@{\hspace{0.5mm}}c@{\hspace{0.5mm}}|c@{\hspace{0.5mm}}|@{\hspace{0.5mm}}c@{\hspace{0.5mm}}|c@{\hspace{0.5mm}}|@{\hspace{0.5mm}}@{}r@{\hspace{0.5mm}}|@{\hspace{0.5mm}}c@{\hspace{0.5mm}}|}

\hline
&\multicolumn{4}{c|}{\textbf{CycleGAN}}&\multicolumn{5}{c|}{\textbf{Pix2PixHD}}\\
 & {\textbf{\FLIP}} & {\textbf{\SURPRISE}} & {\textbf{\NOISE}} & {\textbf{\UNCERTAIN}} & {\textbf{\FLIP}} & {\textbf{\SURPRISE}} & {\textbf{\NOISE}} & {\textbf{\UNCERTAIN}} & {DESIG.} \\
\hline
\textbf{\FLIPCYCLE} & --  & [0.50]  & 0.54  & 0.54  & 0.45  & \textbf{0.64}  & 0.53  & [0.50]  & \textbf{0.59}  \\ \hline
\textbf{\SURPRISECYCLE} & [0.50]  & --  & 0.55  & 0.54  & 0.48  & \textbf{0.63}  & 0.53  & 0.52  & \textbf{0.58}  \\ \hline
\textbf{\NOISECYCLE} & 0.46  & 0.45  & --  & [0.50]  & \textbf{0.40} & \textbf{0.68}  & \textbf{0.42}  & 0.45  & 0.55  \\ \hline
\textbf{\UNCERTAINCYCLE} & 0.46  & 0.46  & [0.50]  & --  & \textbf{0.43}  & \textbf{0.66}  & \textbf{0.43}  & 0.47  & 0.55  \\ \hline
\end{tabular}

\vspace{1mm}
Note: All p-values $< 0.01$, except where `[]' is used to indicate $> 0.1$. Tangible ($\hat{A}_{12}$) differences in bold; for each pair of approaches, the best approach is the one with $\hat{A}_{12} > 0.50$ and its name on the row.
\end{table}

%% file: tables/rq3_retraining_cyclegan.tex
\begin{table*}[t]
\centering
\smaller
\caption{RQ3. $mIoU$ for the DeeplabV3 model retrained using the outputs of the \APPR variants relying on CycleGAN.}
\label{tab:rq3_cyclegan_retraining}
\begin{tabular}{|@{}l|l|l|l|l|l|l|l|}
\hline
\textbf{Retraining set} & \textbf{min} & \textbf{max} & \textbf{median} & \textbf{5th percentile} & \textbf{1st quartile} & \textbf{3rd quartile} & \textbf{Average} \\ \hline


\textbf{\APPRFLIPCYCLE} & 0.00 (+0.00) & \textbf{1.00 (+0.01)} & \textbf{0.51 (+0.07)} & 0.21 (+0.20) & \textbf{0.37 (+0.11)} & 0.67 (-0.10) & \textbf{0.52 (+0.03)} \\ \hline

\textbf{\APPRSURPRISECYCLE} & 0.00 (+0.00) & \textbf{1.00 (+0.01)} & 0.50 (+0.06) & \textbf{0.22 (+0.21)} & 0.36 (+0.10) & 0.67 (-0.10) & \textbf{0.52 (+0.03)} \\ \hline

\textbf{\APPRNOISECYCLE} & 0.00 (+0.00) & \textbf{1.00 (+0.01)} & \textbf{0.51 (+0.07)} & 0.20 (+0.19) & 0.36 (+0.10) & 0.68 (-0.09) & \textbf{0.52 (+0.03)} \\ \hline

\textbf{\APPRUNCERTAINCYCLE} & 0.00 (+0.00) & \textbf{1.00 (+0.01)} & 0.50 (+0.06) & 0.21 (+0.20) & 0.36 (+0.10) & 0.67 (-0.10) & \textbf{0.52 (+0.03)} \\ \hline

\end{tabular}
\vspace{1mm}
\\
Note: The difference with the original DeeplabV3 is shown in parentheses. Best results (per column) are in bold.
\end{table*}

%% file: tables/rq3_retraining_cyclegan_stats.tex
\begin{table*}[t]
\centering
\smaller
\caption{RQ3. p-values and $\hat{E}$ for the data in Table~\ref{tab:rq3_cyclegan_retraining}.}
\label{tab:rq3_cyclegan_retraining_stats}
\resizebox{\textwidth}{!}{%
\begin{tabular}{|l|ll|ll|ll|ll|ll|ll|ll|ll|ll|}
\hline
 & \multicolumn{2}{c|}{\textbf{\APPRFLIPCYCLE}} & \multicolumn{2}{c|}{\textbf{\APPRSURPRISECYCLE}} & \multicolumn{2}{c|}{\textbf{\APPRNOISECYCLE}} & \multicolumn{2}{c|}{\textbf{\APPRUNCERTAINCYCLE}} & \multicolumn{2}{c|}{\textbf{\APPRFLIP}} & \multicolumn{2}{c|}{\textbf{\APPRSURPRISE}} & \multicolumn{2}{c|}{\textbf{\APPRNOISE}} & \multicolumn{2}{c|}{\textbf{\APPRUNCERTAIN}} & \multicolumn{2}{c|}{\textbf{DESIGNATE}} \\
\hline
 & \textbf{$A_{12}$} & \textbf{p-value} & \textbf{$A_{12}$} & \textbf{p-value} & \textbf{$A_{12}$} & \textbf{p-value} & \textbf{$A_{12}$} & \textbf{p-value} & \textbf{$A_{12}$} & \textbf{p-value} & \textbf{$A_{12}$} & \textbf{p-value} & \textbf{$A_{12}$} & \textbf{p-value} & \textbf{$A_{12}$} & \textbf{p-value} & \textbf{$A_{12}$} & \textbf{p-value} \\
\hline
\textbf{\APPRFLIPCYCLE} & & & 0.52& \textbf{2.64E-04}& 0.51& 3.31E-01& 0.51& 1.10E-02& 0.52& \textbf{4.61E-03}& 0.52& \textbf{6.24E-03}& 0.52& \textbf{3.15E-03}& 0.54& \textbf{1.03E-12}& 0.45& \textbf{6.04E-16}\\ \hline
\textbf{\APPRSURPRISECYCLE} & 0.48& \textbf{2.64E-04}& & & 0.49& 2.48E-02& 0.49& 3.92E-02& 0.5& 7.37E-01& 0.51& 3.55E-01& 0.5& 5.69E-01& 0.53& \textbf{1.98E-06}& \textbf{0.44}& 1.53E-27\\ \hline
\textbf{\APPRNOISECYCLE} & 0.49& 3.31E-01& 0.51& 2.48E-02& & & 0.5& 8.08E-01& 0.51& 3.74E-02& 0.53& \textbf{1.10E-05}& 0.51& 2.03E-01& 0.53& \textbf{2.59E-09}& 0.45& \textbf{1.61E-19}\\ \hline
\textbf{\APPRUNCERTAINCYCLE} & 0.49& 1.10E-02& 0.51& 3.92E-02& 0.5& 8.08E-01& & & 0.5& 4.12E-01& 0.51& 9.86E-02& 0.51& 1.89E-01& 0.54& \textbf{2.85E-11}& 0.45& \textbf{8.13E-16}\\ \hline
\end{tabular}%
}
\vspace{1mm} \\
Note: Tangible ($\hat{E}$) and significant ($\mathit{p-value} < 0.01$) differences in bold; for each pair, the best approach is the one with $\hat{E} > 0.50$ and its name on the row.
\end{table*}

%% file: related.tex
\section{Related work}
\label{sec:related}

TACTIC~\cite{zhang2018tactic}, DeepRoad~\cite{zhang2018deeproad}, and DeepTest~\cite{tian2018deeptest} all employ GAN-based perturbations to assess the robustness of DNNs used in autonomous driving. TACTIC applies CycleGAN to simulate diverse weather conditions (rain, fog, snow, night) and evaluates whether self-driving models maintain consistent predictions under environmental changes. Similarly, DeepRoad uses GAN-based transformations to modify road textures and weather conditions, applying metamorphic testing to check for discrepancies in DNN outputs. DeepTest, on the other hand, perturbs images by adjusting brightness, blur, translation, and occlusion effects to identify minor visual modifications that can lead to incorrect predictions. While these methods contribute to DNN robustness evaluation, in our previous work, we demonstrated that a simulator-based search-driven approach (i.e., \DESIGNATE) is more effective for DNN testing and retraining~\cite{attaoui2024designate}. Further, in this paper, we address a complementary problem.




Amini et al. demonstrate that using generative networks to translate simulator images into realistic images reduce the distribution gap between the real-world and synthetic datasets~\cite{Amini2024} thus motivating an approach like \DESIGNATE.
Further, they developed SAEVAE, a generative network that performs better than CycleGAN in terms of reduction of simulator-to-real gap, training, and prediction time. 
\APPR is modular and the CycleGAN component can be replaced by any better-performing GAN, including SAEVAE.

Baresi et al. leverage diffusion models to test vision-based DNNs~\cite{Baresi2025}; however, they do not fully leverage the input generation capabilities of these models because, to avoid the problem of deriving a ground truth for the generated images, they use diffusion models to replace  elements (e.g., background) that shall not affect the DNN outcome. Recently, they extended the approach by automatically identifying the items to replace by processing the textual descriptions automatically generated by LLMs~\cite{Baresi2025b}.
The accuracy metrics proposed in this paper may enable to better leverage the generative power of diffusion models (e.g., to generate images from scratch).


%% file: conclusion.tex
\section{Conclusion}
\label{sec:conclusion}

We proposed an approach to address the oracle problem when performing search-based, simulator-driven testing of Deep Neural Network (DNN) components for computer vision. The oracle problem arises when the simulator cannot generate ground truth data (e.g., space simulators do not produce segmentation maps). Test input images generation is performed by combining meta-heuristic search algorithms, Generative Adversarial Networks, and simulators that cannot generate ground-truth data. Further, we propose four fitness functions inspired by related work on DNN testing; they leverage geometric transformations, noise perturbation, surprise adequacy, and uncertainty estimation. 

Our contributions are an approach to perform simulator-based search-driven testing of DNN components in absence of ground truth, an empirical assessment of strategies to drive test input generation in absence of ground truth, an empirical assessment of the impact of ground truth data on GANs for the generation of realistic inputs.

Our evaluation, conducted with a DNN for Mars exploration, compared the effectiveness of the proposed fitnesses with a fitness based on ground-truth data, in a controlled experiment where the ground truth is used for the generation of realistic images with GANs. Further, we assessed how relying on a GAN that does not leverage ground-truth data harms the effectiveness of the approach.

Our results demonstrate that leveraging geometric transformations and surprise adequacy lead to best results for both testing and retraining tasks. A fitness based on geometric transformations identify inputs leading to minimal DNN performance while surprise adequacy maximizes diversity. Both these two fitnesses help improving DNN accuracy although they, expectedly, do not perform like an approach leveraging the ground truth. Further, we demonstrated that the use of GANs that do not leverage ground-truth data does not negatively affect the effectiveness of the approach. However, the fitness based on geometric transformations is the only one that both maximizes test diversity and achieves a retraining accuracy similar to that of an approach relying on ground truth. 

Future work will explore how the proposed fitness functions can be applied to oracle-free testing approaches that use generative AI models instead of simulators.


